\documentstyle[aps,epsf,floats,twocolumn,prb,amssymb]{revtex}
{\bf }

\begin{document}
\twocolumn[\hsize\textwidth\columnwidth\hsize\csname@twocolumnfalse%
\endcsname
%\draft

\title{Griffiths Effects in Random Heisenberg Antiferromagnetic $S=1$ Chains}
\author{Kedar Damle}
\address{Physics Department, Harvard University, Cambridge, MA 02138
}
\date{January 7, 2002}
\maketitle

\begin{abstract}
{I consider the effects of enforced dimerization on random Heisenberg
antiferromagnetic $S=1$ chains.
I argue for the existence of novel Griffiths phases
characterized by {\em two independent dynamical exponents}
that vary continuously
in these phases; one of the exponents
controls the density of spin-1/2 degrees of freedom in the low-energy
effective Hamiltonian, while the other controls the corresponding density
of spin-1 degrees of freedom. Moreover, in one of these Griffiths phases,
the system has very different low
temperature behavior in two different parts of the phase which are separated
from each other by a sharply defined crossover line; on one side
of this crossover line, the system `looks' like a $S=1$ chain
at low energies, while on the other side, it is best thought
of as a $S=1/2$ chain. A
strong-disorder RG analysis makes it possible to analytically obtain
detailed information about the low temperature behavior of physical
observables such as
the susceptibility and the specific heat, as well as identify an experimentally
accessible signature of this novel crossover.}
\end{abstract}

%\pacs{PACS numbers: 75.10.Jm, 78.70.Nx, 75.50.Ee, 71.30.+h}
\vskip 0.3 truein
]

\section{Introduction}
Quantum spin chain systems are known to exhibit many interesting states of
matter at low temperature that arise from the fascinating interplay
between quantum fluctuations, correlation effects, and the effects of
quenched disorder.
They are particularly interesting
from a theoretical point of view, as they provide experimentally
realizable examples in which this interplay, common to many other
condensed-matter systems, can be studied in detail.

Among the most studied such systems is the Heisenberg antiferromagnetic
(HAF) $S=1$ chain (for an experimental review see Ref.~\onlinecite{Expt1})
with Hamiltonian:
\begin{equation}
{\cal H} = J\sum_{i} \hat{S}_i \cdot \hat{S}_{i+1} \; ,
\label{1HAF}
\end{equation}
where $J$ is positive, and the operators $\hat{S}$ represent the spin angular
momentum of spin-1 objects on site $i$ of a one-dimensional lattice.
As Haldane\cite{Hal1} showed many years ago,
quantum fluctuations in the $S=1$ HAF chain are strong
enough to rule out
even quasi-long range antiferromagnetic order of the type
present in $S=1/2$ chains, and
the system is in the so-called `Haldane' phase characterized
by a singlet ground-state and gap to all bulk excitations. However,
this phase possesses a kind of `topological' order\cite{Top} that results
in the presence of almost free sub-gap boundary spin-1/2 degrees of
freedom in a large but finite system of length $L$
with free ends; these spin-1/2s are coupled to each other by an
effective exchange coupling $J_{\rm eff} \sim (-1)^{L+1}e^{-cL}$
(with $L$ in lattice units, and $c$ proportional to the inverse
ground-state correlation length), and will play a crucial role in much of
the analysis in this paper.

A very intuitive and useful picture of the Haldane phase is the
`valence-bond' solid\cite{AKLT} description:
The basic idea is to think of each spin-1 as a (symmetrized) pair of 
spin-1/2 objects; each site thus has two `indistinguishable' spin-1/2
objects on it.
A good description of the ground state is then
obtained by pairing one of the spin-1/2 at any site
into a singlet state with a spin-1/2 on its neighbor to the right, and
pairing the other spin-1/2
into a singlet state with a spin-1/2 on its neighbor to the left.
This gives a state with one singlet bond across each link of
the lattice, i.e the (1,1) state. The topological
order present in the Haldane phase can now be thought
of simply as the ability to `walk' across the length of the system along
singlet bonds without encountering any breaks, and the unpaired spin-1/2s
at each end of a chain with free boundaries model the sub-gap
end-states characteristic of the Haldane phase.

The stability of this phase to various experimentally relevant
perturbations has also been investigated.
For instance, it is known\cite{AffHal} that
the Haldane phase is stable to weak
enforced dimerization $\delta$ in the exchange
couplings ($\delta$ parametrizes the extent to which the
even bonds are stronger than the odd bonds), and there are no qualitative differences in the
low-energy properties of the system until $|\delta|$ exceeds a critical value
$\delta_c$. Beyond this point, the ground state changes
character, and the system enters a different gapped state without the
topological order or the subgap spin-1/2s on free ends.
For positive $\delta >\delta_c$, this dimerized phase
corresponds to the (2,0) valence-bond state in which both
spin-1/2 degrees of freedom at any even-numbered physical site form
singlets with
spin-1/2s on its neighbor to its right, and the number of singlet bonds across
each link of the lattice alternates between two and zero (similarly, the
(0,2) state is favored by the system when $\delta < -\delta_c$).

The effects of quenched randomness $R$ in
the exchange couplings (with
all $J_i$ still positive) have also been studied\cite{HymYan,MonJolGol}
{\em at} $\delta = 0$ (in the
presence of randomness, $\delta$ is defined in terms of the
average values for even and odd bonds). The topological order
characteristic of the Haldane state persists for $R$ less than a critical
value $R_c$, and
the Haldane phase is thus stable to weak randomness. On the other hand,
a different, randomness dominated
state is stabilized for $R>R_c$. In this `Spin-1 Random Singlet'
(RS$_{\rm 1}$) state, the interplay of disorder and quantum mechanics
locks each spin-1 into a singlet state with some other spin-1; the two spins
in a given singlet pair can have arbitrarily large spatial separation, with
disorder determining the actual choice of partner for each spin;\cite{DSF1}
in the valence bond picture, this large $R$ state is thus characterized by
a particular {\em random} pattern of {\em double} bonds.\narrowtext
\begin{figure}
\epsfxsize=\columnwidth
\centerline{\epsffile{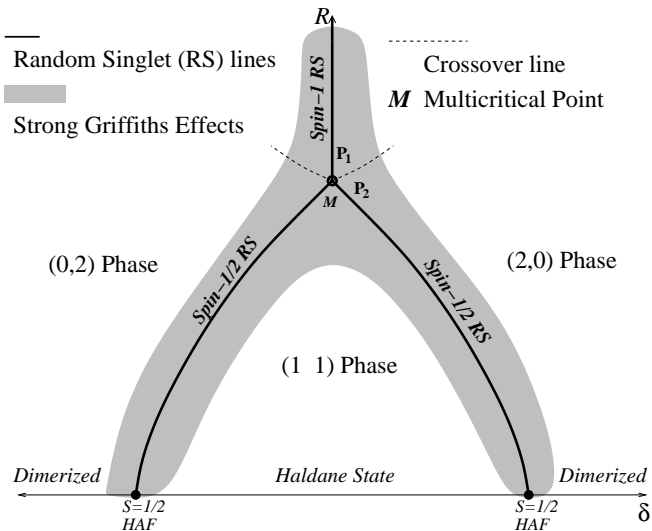}}
\vspace{0.15in}
\caption{Schematic phase diagram as a function of
randomess ($R$) and dimerization ($\delta$). Details in text
}
\label{phased}
\end{figure}

Although the topological order survives for all $R<R_c$, it turns out
that disorder has
a dramatic effect on the low-energy properties near the transition,
resulting in arbitrarily low-energy
excitations with a power-law density of states controlled
by a single non-universal (R dependent) {\em dynamical exponent} $z$.
These low-energy excitations are associated
with large anomalous regions of the sample in which
the couplings locally favor a pattern of singlet bonds more
characteristic of the other `nearby' phase(s) (see below). Such
`Griffiths effects', whereby rare disorder-induced fluctuations
in the interactions over extended regions of space
give rise to
{\it non-universal} singular contributions that the dominate low-energy
properties, are among the more interesting and ubiquitous aspects of the
physics of low-dimensional random quantum systems, and below,
I focus on precisely this physics for the general $\delta \neq 0$
case.

\section{Physical Picture and Motivation}
\label{motiv}
Let us start with
the overall structure of the
phase diagram in the ($R$,$\delta$) plane:
To begin with, note that
the nature of the ground state will
not change qualitatively from that at $\delta = 0$ as long as
$R<R_c$ and $\delta$ is small enough.
On the other hand, by analogy with the corresponding situation
in random HAF $S=1/2$ chains,\cite{HyYaBhGi} the RS$_1$
state for $R>R_c$ will be unstable to adding a tiny amount of dimerization.
For small $\delta >0$, the resulting pattern of singlet bonds in the
ground state
will be the random analog of the (2,0) state of the pure system, and
topologically indistinguishable from the latter. Likewise, the state for
infinitesimal $\delta < 0$ will be topologically identical to
the (0,2) state. Putting all this together, we are led to a phase
diagram that looks something like the one sketched in Fig~\ref{phased}.
The critical
point\cite{HymYan,MonJolGol} at $R= R_c,\delta = 0$ is thus
really a {\em multicritical point} at which the (2,0), (0,2) and
(1,1) phases meet, while the RS$_1$ line is seen to be the critical
phase boundary between the (0,2) and (2,0) phases. 

To complete this part of our discussion, we need to identify the universality class of the
phase-boundary between the (1,1) phase and either of the dimerized phases.
To this end, note that the physics asymptotically close to $\pm \delta_c$ at $R=0$ is that of a {\em spin-$1/2$}
chain with a small value of enforced dimerization: heuristically,
this may be understood
by first putting down a single valence bond on each even link (for
$\delta > 0$) to
partially `screen' out the imbalance in the odd and even exchange couplings;
this leaves behind
a $S=1/2$ HAF chain with $\delta_{\rm eff} \sim \delta - \delta_c$.
Now, we know\cite{DSF1} that adding disorder to the $S=1/2$ HAF at $\delta_{\rm eff} = 0$ 
leads to a spin-1/2 Random Singlet (RS$_{1/2}$) state
(with a random pattern of {\em single} valence bonds statistically identical
to the pattern of {\em double} bonds in the RS$_1$ phase), and thus,
the $\delta \neq 0$ phase boundaries between the (1,1) state
and the (2,0) or (0,2) states are lines along which 
the system is in the RS$_{1/2}$ critical state.

\begin{figure}
\epsfxsize=\columnwidth
\centerline{\epsffile{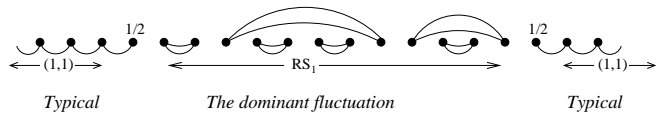}}
\vspace{0.15in}
\caption{A cartoon for Griffiths effects in
the Gapless Haldane phase at $\delta = 0$
}
\label{griffarg1}
\end{figure}

Of course, the topological labels of the
different phases and the universality classes of various transitions are not
the whole story, and we need to consider the role of Griffiths effects
in various
regimes to really understand the low-energy behavior of the phases.
Consider, for starters, the Gapless Haldane regime at $\delta=0$.
The gapless spectrum obtained in
Refs~\onlinecite{HymYan,MonJolGol} can be understood by thinking about the effect
of a single rare region of length $L$ in which the exchange couplings are such
that the preferred pattern of singlet bonds in this region is
more characteristic of a system at a nearby point in either the random
(2,0) or (0,2) phases, or on the critical RS$_1$ line separating them
(outside of this
rare segment of length $L$, the system is in the `typical' (1,1)
state). In all such cases, the central segment (see Fig~\ref{griffarg1})
can be thought of as
an `insulating' barrier that separates two essentially semi-infinite chains in
the Haldane phase, implying the presence of low-energy spin-1/2 degrees
of freedom localized at the ends of the (1,1) segments.\narrowtext
These spin-1/2s are
coupled to each other across the `barrier' by an antiferromagnetic exchange
coupling of order $e^{-c_1L}$, while the probability for such an anomalous
disorder
configuration to occur is also exponentially small: $p_L \sim e^{-c_2L}$.
Averaging over different possibilities, with anomalous regions of
varying length $L \gtrsim -\ln(\Omega)/c_1$, then yields a power-law for the density of spin-1/2
degrees of freedom that model the physics below the energy scale $\Omega$:
$n_{1/2} \sim \Omega^{1/z}$, with
a non-universal exponent $z$ (here, and
henceforth, $\Omega$ denotes some suitable low energy scale much
smaller than the microscopic scale $J_{\rm typ}$). Now, if two such anomalous regions
occur essentially `next' to each other ({\em i.e} separated by a segment with
typical exchange couplings of length $L \lesssim -\ln(\Omega)$), and if
$L$ is {\em even},
then the end spin-1/2 degrees of freedom of this intervening typical
segment will bind
ferromagnetically to each other producing an effective spin-1 as
far as the low-energy dynamics below the cutoff scale $\Omega$ is concerned.
The probability for this to happen is dominated by the requirement that
two anomalous regions, that are usually separated by lengths of order
$\Omega^{-1/z}$, occur essentially next to each other---this immediately
implies that the density of spin-1 degrees of freedom in the effective
Hamiltonian will scale as the square of the density of spin-1/2
objects, consistent (apart from a log-correction) with the RG results
of Ref~\onlinecite{HymYan}.

\begin{figure}
\epsfxsize=\columnwidth
\centerline{\epsffile{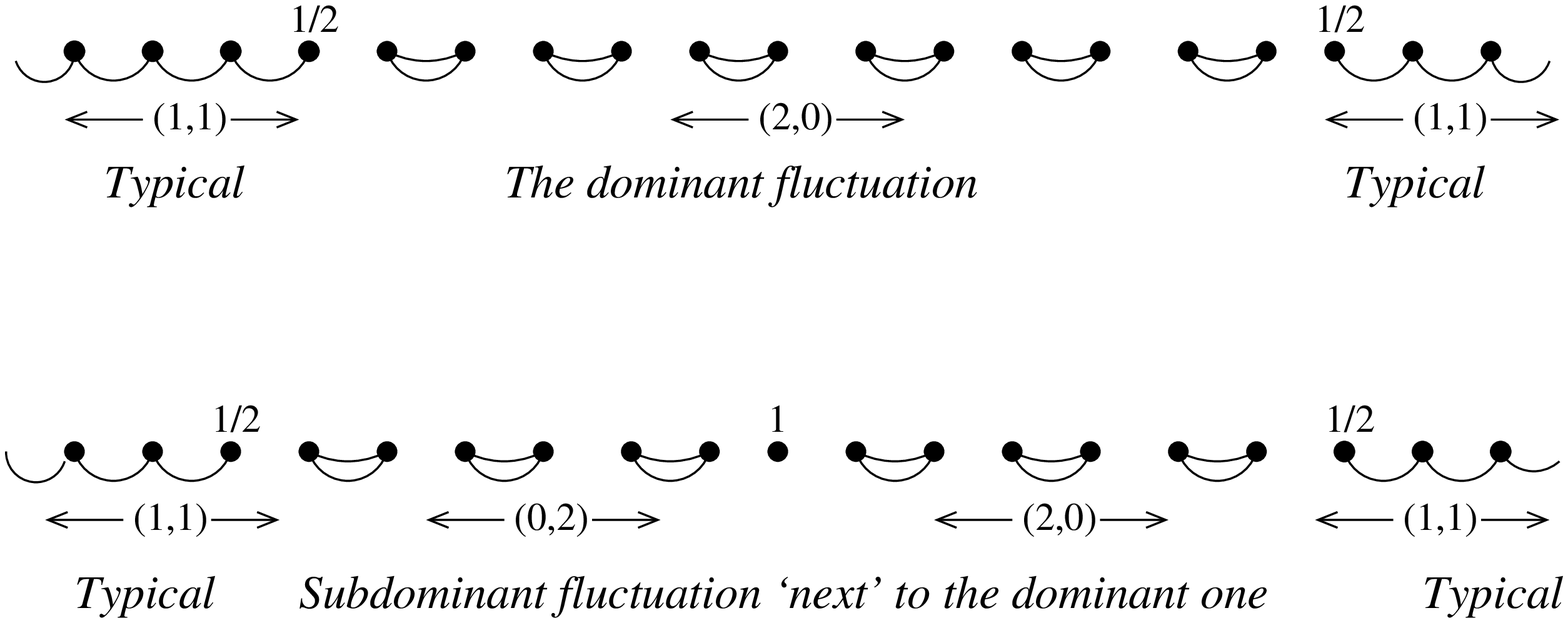}}
\vspace{0.15in}
\caption{A caricature of Griffiths effects in the (1,1) phase at positive $\delta$
}
\label{griffarg2}
\end{figure}

It turns out that similar heuristic arguments can be employed to qualitatively
understand Griffiths effects at $\delta > 0$ in the (1,1) phase, {\em i.e},
closer to the phase boundary to the (2,0) state (see fig~\ref{griffarg2}):
In this case, the dominant disorder-induced fluctuation will consist
of large regions of the sample that locally want to be in the (2,0)
state---the probability of their occurrence will control the
exponent $z_{1/2}$ that determines the density $n_{1/2} \sim \Omega^{1/z_{1/2}}$ of low-energy spin-1/2
degrees of freedom below scale $\Omega$,
while much rarer regions that locally prefer the
(0,2) phase or the RS$_1$ critical state will only provide
sub-dominant corrections as far as the value of $z_{1/2}$ is concerned.\narrowtext
However, it is clearly impossible (simply for parity reasons) to have
two anomalous segments in the (2,0) phase separated from each other
by an {\em even}
segment---one of these anomalous regions necessarily has to be of a
much rarer sub-dominant type (either a (0,2) region, or
one with a spin-1 random singlet pattern of valence bonds)
in such a situation. These are
precisely the configurations that behave at low energies as a
spin-1 object, and thus, the spatial density of spin-1 degrees of freedom
below scale $\Omega$ vanishes faster than $\Omega^{2/z_{1/2}}$.
Introducing a second
exponent $z_{1} \leq z_{1/2}/2$ to describe this behavior,
we therefore arrive at
$n_{1} \sim \Omega^{1/z_1}$. Moreover, it is also clear from this
argument that
$z_{1/2}$ will diverge as we approach the transition at non-zero $\delta$
into either the (2,0) or (0,2) phases (after all, what was
the dominant rare fluctuation becomes the typical configuration as one
crosses the transition!), while $z_1$, whose value near the RS$_{1/2}$
phase boundary is primarily determined by the
{\em sub-dominant} fluctuation, will
remain finite at this transition.

In this analysis of the (1,1) phase,
I have been unavoidably led to a description
in terms of {\em two independent} dynamical exponents. This is, of course,
formally quite interesting---however, since $z_1 \leq z_{1/2}/2$ throughout
this phase, this additional `tuning-knob' does not lead to
any qualitative changes in the low-energy physics as we move around in
this phase. On the other had, the situation in the (2,0) (or (0,2)) phase in
the vicinity of
its phase boundaries is dramatically different, and this is what I
turn to next.

\begin{figure}
\epsfxsize=\columnwidth
\centerline{\epsffile{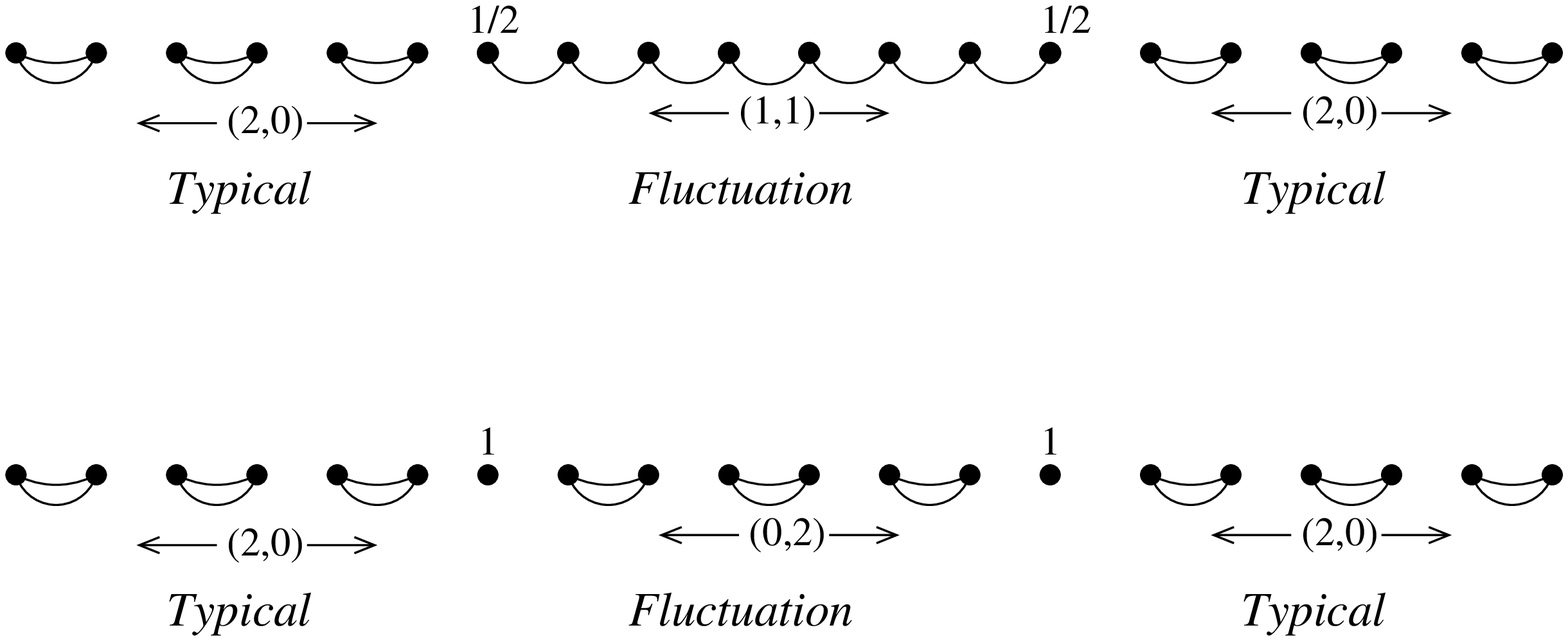}}
\vspace{0.15in}
\caption{Cartoon for Griffiths effects in the (2,0) phase
}
\label{griffarg3}
\end{figure}

To begin with, let us focus on two representative
points in this phase (see Fig~\ref{phased}),
lying on an arc drawn around the multicritical point
and going from the RS$_1$ critical line to the RS$_{1/2}$ critical line.
Consider first the point $P_1$ very close to
the RS$_1$ line: The low-energy dynamics at this point will be dominated
by rare large regions that locally want to be at a nearby point in
the (0,2) phase, embedded in a more typical background which is in
the (2,0) phase (see fig~\ref{griffarg3}). Such an anomalous region clearly has residual spin-1 objects
at either end, and the low-energy dynamics below scale $\Omega$
will be dominated by a density $n_1 \sim \Omega^{1/z_1}$ of these, with
$z_1$ being controlled by the probability for such anomalies to occur at
the corresponding length scales.\narrowtext
Of course, there will also be sub-dominant disorder fluctuations that involve
much rarer regions locally in the (1,1) phase; these have spin-1/2
degrees of freedom at their ends, and the low-energy dynamics will therefore
also have sub-dominant contributions from a density
$n_{1/2} \sim \Omega^{1/z_{1/2}}$
of these (with $z_{1/2} \ll z_{1}$ being controlled by the probability
of occurrence for the (1,1) anomalies). Simply put, in this regime of the
(2,0) phase, the low-energy behavior
is more or less that of a spin-1 system.
Next, consider the point $P_2$ chosen very close to the RS$_{1/2}$
phase boundary.
The situation is now completely reversed, and the dominant Griffiths
effects arise from rare regions locally at a nearby point in the (1,1)
phase. Thus, $z_{1/2}$ will be much larger than $z_1$ in this regime,
and the low-energy behavior is predominantly that of a system made up
of spin-1/2 objects.

Clearly, as we move along the arc from $P_1$ to
$P_2$, the low-energy response will smoothly interpolate between these two
extremes. Thus, the system in either the (2,0) or (0,2) phase has a curious
ability to look {\em qualitatively different at low energies} in different parts of the
phase, undergoing a metamorphosis from
a spin-1/2 system into a spin-1 system as we move around in the phase.
In particular, one expects that there will be some
intermediate point along the arc at which $z_1 = z_{1/2}$; at this
point, {\em both} spin-1/2 and spin-1 degrees of freedom will play
a role in determining the low-energy behavior of the system.
[More generally, one expects a crossover line emanating from the
multicritical point, along which $n_{1/2} \sim n_1 \sim \Omega^{1/z}$ as
$\Omega \rightarrow 0$.]

These heuristic considerations clearly demonstrate that there
is enough new and interesting physics at $\delta \neq 0$ to warrant a
more serious analysis, and this is what I turn to next.

\section{Effective model and the RG approach}
To go beyond the qualitative arguments of the previous section, one needs to
analyze an appropriate effective model in a controlled manner.
To fix the form of this effective
model, it is useful to specialize to
a situation in which the bonds $J_{i}$ in our Hamiltonian
\begin{eqnarray}
{\cal H} & = & \sum_{i} J_i \hat{S}_i \cdot \hat{S}_{i+1}
\end{eqnarray}
only take values $1$, $\epsilon_s >0$,
or $\epsilon_w > 0$ ($s$ and $w$ stand for `strong' and `weak', with
$1 > \epsilon_s > \epsilon_w$).
For $i$ even, $J_i$ is $1$ with probability $1-p_s$ (the $J_i$ for $i$
even are thus `strong'),
and $\epsilon_s$ with probability $p_s$, while, for $i$ odd, $J_i$
is $1$ with probability $1-p_w$ (the $J_i$ for odd $i$
are thus `weak'), and $\epsilon_w$
with probability $p_w$ (with $p_w > p_s$).
For concreteness, it is also useful to
take the $p_{s/w}$ and $\epsilon_{s/w}$ all much smaller than $1$.
In this case, it is appropriate to think in
terms of a collection of fairly large segments of a pure $S=1$ HAF chain in
the Haldane state, coupled to each other by the $\epsilon$-bonds, and
described in terms of pairs of the sub-gap boundary spin-1/2 degrees
of freedom at their ends.

Given that $p_w > p_s$, a typical segment of this kind will be, more likely
than not, flanked on {\em both} sides
by {\em $\epsilon_w$-bonds}. Consequently, such a segment
will be {\em odd in length}, and its low-energy behavior will be described by
a pair of spin-1/2s
coupled to each other by an `odd' bond which is {\em antiferromagnetic},
and drawn from a distribution that is calculable in terms of the
length-distribution of the pure segments (the designation `odd' has nothing
to do with the parity of the segment length, and is
chosen to conform to the notation of Ref.~\onlinecite{HymYan} when
$\delta = 0$)---the bonds connecting this `odd' pair to the rest of
the system on either side are `even'
bonds of the {\em $w$-type} (the designation `even' is again chosen
to conform with Ref~\onlinecite{HymYan} when $\delta = 0$; clearly
all `even' bonds are always antiferromagnetic).
On the other hand, a segment flanked on one side by an $\epsilon_s$-bond,
and on the other by a $\epsilon_w$ bond, will be {\em even
in length}, and its low-energy description will consist of a pair of
spin-1/2s coupled by
a {\em ferromagnetic} `odd' bond (whose {\em modulus} has the same
distribution as in
the earlier antiferromagnetic case). Of course, this odd ferromagnetic
pair is flanked by
one {\em $s$-type} even bond and another {\em $w$-type} even bond;
odd ferromagnetic bonds thus have a {\em $s$-flank} and a {\em $w$-flank}.\narrowtext
\begin{figure}
\epsfxsize=\columnwidth
\centerline{\epsffile{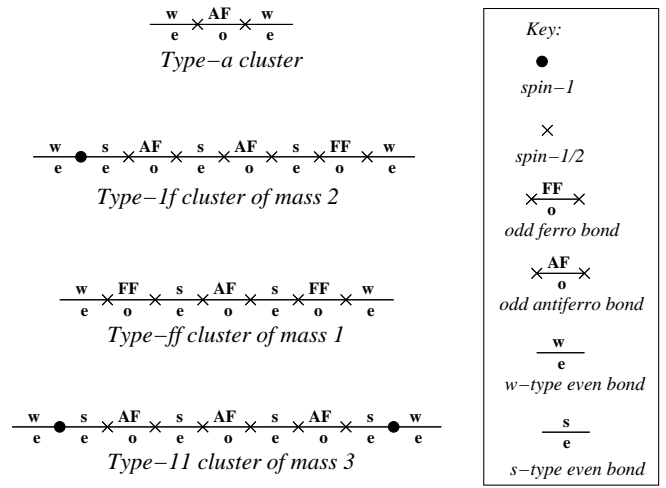}}
\vspace{0.15in}
\caption{Some examples of the different types of clusters that make
up our effective model.
}
\label{effmodel}
\end{figure}

Moreover, and this is
crucial, the other $\epsilon$-bonds in the vicinity of any such segment
(that contributes a ferromagnetically coupled odd pair of spin-1/2s
to the effective model)
are again
{\em more likely to be of the $w$-type}, and thus, there is an
enhanced probability for finding another even-length pure segment
(contributing {\em another} pair of spin-1/2 coupled by a
{\em ferromagnetic} odd
bond) close-by on the {\em $s$-flank} of this segment. In other words,
a ferromagnetic odd bond in the effective theory
likes to have another ferromagnetic odd bond (or a spin-1 object; see below)
close-by on its $s$-flank.
Likewise, an odd-length segment flanked on {\em both} sides
by $\epsilon_s$ bonds (described at low-energies by an odd
antiferromagnetic pair of spin-1/2s flanked on both sides by $s$-type even
bonds) is more likely than not
to have nearby even-length segments in the Haldane phase; thus, in the
effective theory,
an odd antiferromagnetic bond flanked by {\em $s$-type} even bonds on
{\em both} sides
side likes to have odd ferromagnetic bonds (or spin-1s; see below)
close by on {\em both} sides.
Finally, two $\epsilon$ bonds right next to each other will give
rise to a spin-1 object connected to the rest of the system by
a $w$-type even bond on one side, and a $s$-type even bond on the other
(obviously, such a spin-1 object also has a tendency to
have another ferromagnetic odd bond (or spin-1) close-by on its $s$-flank).
Clearly, one could equally well represent this spin-1 by
two spin-1/2s connected by a ferromagnetic odd bond with strength much
bigger than the cutoff,\cite{HymYan,MonJolGol}
and in this sense, this is just a special case
of a ferromagnetically coupled odd-pair---however, I prefer to explicitly
introduce spin-1 objects into my description here.

This dimerization induced clustering tendency of the
ferromagnetically coupled odd pairs (or spin-1s) in the effective model
immediately implies that individual pairs of spin-1/2s coupled
by odd bonds (and their special case: spin-1s)
are not the elementary constituents of our
effective theory (this should be contrasted with the $\delta = 0$ case,
where the low-energy effective model does consist of statistically independent
odd pairs coupled together by even bonds).\cite{HymYan} Clearly, the solution
is to think not in terms of individual odd pairs, but
in terms of {\em clusters}: The simplest cluster, the $a$-type cluster,
is just a pair
of spin-1/2s coupled by an {\em antiferromagnetic} odd bond, and
flanked on both sides by {\em $w$-type even bonds}. Any other cluster
of {\em mass} $\mu$ is made up of a string of $\mu$ antiferromagnetically coupled odd pairs of spin-1/2s ($\mu = 0 \, , \, 1 \, , \, 2 \, , \, 3 \dots $), each flanked by $s$-type even bonds on both sides
(the total number of $s$-type even bonds is thus $\mu+1$),
terminating at each end in a ferromagnetically coupled odd pair of spin-1/2s.
Of course, either,
or both, of the terminating pairs could equally well be replaced by a
single spin-1 each, and
there are thus three kinds of clusters: $11$
(with two spin-1s at the two ends),
$1f$ (with a spin-1 at one end and a pair of spin-1/2s coupled with an
odd ferromagnetic bond at the other end), and $ff$ (with ferromagnetically
coupled odd pairs of spin-1/2s at both ends).

Our effective model\cite{foot0} is thus made up
of these four types of statistically independent clusters,
connected to each other by intervening $w$-type even bonds. Furthermore,
the absolute values of all
odd bonds, regardless of their sign, are expected to be governed by a
single distribution $P_o$. On the
other hand, the $w$-type and $s$-type even bonds will have 
two different distributions, which we denote by $P_w$ and $P_s$ respectively.
Furthermore, since each end of any cluster is expected to
independently be either a spin-1, or a ferromagnetically coupled odd
pair of spin-1/2s, the relative abundances of the
$11$, $1f$, and $ff$ clusters can be parametrized by a single
probability $g$ for an end to be a spin-1 (with the probability
for an end to be a ferromagnetically coupled pair of spin-1/2s being
$1-g$). Similarly, the density of the `elementary' $a$-type clusters
relative to the density of all other clusters can be parametrized
by a probability $f_a$ for any given cluster to be of type $a$.
Finally, since we are dealing with uncorrelated disorder, it is
reasonable to expect that the masses $\mu$ of the
$11$, $1f$, and $ff$ clusters are all characterized by a single exponential
distribution, with probability $\propto q^\mu$ for any cluster to have
mass $\mu$.

All these expectations are borne out by my detailed calculations below;
for now,
I only note that the effective model is completely specified by the three
probabilities $g$, $f_a$ and $q$, the three probability distributions
$P_s$, $P_w$ and $P_o$, and the number of
$w$-type even bonds $N_w$. Naturally, the initial values of these parameters
and distributions (at the intermediate
energy scale $\Omega_0$ below which this model becomes applicable)
are determined by the complicated interplay of dimerization and randomness in
the microscopic model---fortunately, their precise values and
functional forms will, for the most part, be unimportant as
far as the physics at scales
$\Omega \ll \Omega_0$ is concerned.

To get at this low-energy physics,
it is convenient to use a strong-disorder RG approach\cite{HymYan,DSF1,MaDaHu}
in which we iteratively diagonalize the most strongly coupled parts
of the Hamiltonian: 
Thus, at each step, 
we focus on the strongest bond $|J_{\rm max}| \equiv \Omega$
in the system. Consider, first, the case when this is an
antiferromagnetic bond:
If this bond connects two spin-1/2s
(such a bond can clearly be either even or odd),
we freeze the two spin-1/2s into a singlet state, and
introduce a renormalized coupling between the neighboring spins on either
side: ${\tilde{J}}=
J_LJ_R/J_{\rm max}$; here $J_L$ and $J_R$ are the bonds
immediately adjacent to $J_{\rm max}$, respectively to its left and
to its right. If this bond connects two spin-1s (in this case, $J_{\rm max}$
must necessarily be an even bond), we again freeze
them into their singlet ground state, and couple the neighboring spins on
either side to each other with the renormalized bond ${\tilde{J}}=
J_LJ_R/J_{\rm max}$. Finally, consider the case where $J_{\rm max}$ couples
a spin-1 with a spin-1/2 (again, in this case, $J_{\rm max}$ must be even),
with this spin-1/2 coupled to its other
neighbor with a bond $J_{\rm half}$, and this spin-1 coupled to its
other neighbor with a bond $J_{\rm one}$. In this case, we form
a renormalized spin-1/2 object representing the doublet ground state of
this pair. The neighboring bonds $J_{\rm half}$ and $J_{\rm one}$
remain unchanged in magnitude after the RG step, but $J_{\rm half}$
{\em changes sign}.
Finally, if $J_{\rm max}$ is ferromagnetic (in this case, $J_{\rm max}$ has
to be an odd bond connecting two spin-1/2s), we
put this pair of spin-1/2s into their triplet ground state, and replace
it by an equivalent spin-1.

This formulation of the RG is completely equivalent to the one used in
Ref~\onlinecite{HymYan}, and ignores $O(1)$ coefficients in the RG
recursion relations as well as the distinction between the bond $J$ coupling
a pair, and the gap in its spectrum---as in that case, I expect it to give
accurate results when the effective value of disorder in the low-energy theory
is large. My calculations will therefore be essentially exact
in the regime of strong Griffiths effects (with a large value for
at least one of the dynamical exponents), and are expected to be accurate
even for smaller values of the dynamical exponents.\cite{Igl1}
In the next two sections, I use this RG approach to analyze the
low-energy physics of the effective model.

\section{RG flow-equations and fixed-points}
It is straightforward, if a little tedious, to write down
the flow equations that describe the iterated action of the RG rules on our
effective model, and
this is what I turn to next.

But first, some additional notation:
Let us introduce $N_{ff}(\mu)$, $N_{1f}(\mu)$, $N_{11}(\mu)$, and $N_a$ to
denote the number of mass-$\mu$ $ff$, $1f$, $11$ clusters, and
$a$-type clusters respectively.
Thus, we have
\begin{eqnarray}
N_{ff}(\mu) & = & (1-f_a)(1-g)^2(1-q)q^\mu N_w \; , \nonumber \\
N_{1f}(\mu) & = & 2(1-f_a)g(1-g)(1-q)q^\mu N_w \; , \nonumber \\
N_{11}(\mu) & = & (1-f_a)g^2(1-q)q^\mu N_w \; , \nonumber \\
N_a & = & f_a N_w \; .
\label{ansatz}
\end{eqnarray}
In addition, let us denote the total number of $ff$, $1f$, and $11$
clusters by $N_{ff}$, $N_{1f}$, and $N_{11}$ respectively.
It is also convenient to introduce the log-cutoff $\Gamma = \ln(\Omega_0/\Omega)$, and the log-couplings $\beta = \ln(\Omega/|J|)$, and keep
track of the probability distributions of bond-strengths in terms of
the probability distributions $P_{s/o/w}(\beta|\Gamma)$ of the corresponding
log-couplings. And finally, let us introduce $P_{s/o/w}^{0}(\Gamma)$ 
to refer to $P_{s/w/o}(0|\Gamma)$ respectively.

The flow equations for the probabilities read
\begin{eqnarray}
\frac{df_a}{d\Gamma} & =  & \!\!(1-f_a)[P_s^0(1-q)(1-g^2(1-f_a))-f_a(P_o^0+P_w^0)]
\; , \nonumber \\
\frac{dq}{d\Gamma} & =  &\!\! (1-q)[P_w^0(1-f_a)(1-g^2(1-q))-q(P_o^0+P_s^0)]
\; , \nonumber \\
\frac{dg}{d\Gamma} & =  & (1-g)P_o^0 - g(P_s^0q+P_w^0f_a),\!\!\!\!\!\!\!\!\!\!\! \!\!\!\!\!\!\!\!\!\!\!\!\!\!\!\!
\label{fqgeqn1}
\end{eqnarray}
while those for the functions $P_{s/w/o}$ read:
\begin{eqnarray}
\frac{\partial P_s}{\partial \Gamma} & = & \frac{\partial P_s}{\partial \beta}
+ P_s^0P_s +\nonumber \\
&& \!\!\!\!\!\!\!\!\!\!\!\!\!\!\!\!\! +
(qP_o^0+(1-q)(1-f_a)g^2P_w^0)(P_s \otimes P_s -P_s), \nonumber \\
\frac{\partial P_w}{\partial \Gamma} & = & \frac{\partial P_w}{\partial \beta}
+ P_w^0P_w +\nonumber \\
&& \!\!\!\!\!\!\!\!\!\!\!\!\!\!\!\!\! +
(f_aP_o^0+(1-q)(1-f_a)g^2P_s^0)(P_w \otimes P_w -P_w), \nonumber \\
\frac{\partial P_o}{\partial \Gamma} & = & \frac{\partial P_o}{\partial \beta}
+ P_o^0P_o +\nonumber \\
&& \!\!\!\!\!\!\!\!\!\!\!\!\!\!\!\!\! +
(AP_s^0+BP_w^0)(P_o \otimes P_o -P_o), \!\!\!\!\!\!\!\!\!\!\!\!\!\!\!\!\!\!
\label{Pswoeqn1}
\end{eqnarray}
where $P \otimes P$ stands for the convolution
\begin{eqnarray}
P \otimes P & = &\int \limits_{0}^{\infty}\int \limits_{0}^{\infty}
d\beta_1 d\beta_2 P(\beta_1|\Gamma)
P(\beta_2|\Gamma)\delta(\beta-\beta_1-\beta_2), \nonumber
\end{eqnarray}
and $A(\Gamma)$ and $B(\Gamma)$ are defined as
\begin{eqnarray}
A(\Gamma) & = & \frac{(1-f_a)(1-g(1-q))^2}{2(1-g)+(2g-1)(f_a+q)-2gqf_a},
\nonumber \\   
B(\Gamma) & = & \frac{(1-q)(1-g(1-f_a))^2}{2(1-g)+(2g-1)(f_a+q)-2gqf_a}.
\nonumber 
\end{eqnarray}
Finally, the last equation governs the $\Gamma$ dependence of $N_w$:
\begin{eqnarray}
\frac{dN_w}{d\Gamma} & = & -[P_w^0 +(1-q)(1-f_a)g^2P_s^0 +f_aP_o^0]N_w.
\label{Nweqn1}
\end{eqnarray}

Some comments are in order before we go any further with our analysis:
First of all, the very fact that these equations can be written down
at all relies on the fact that of our ansatz, Eqn~(\ref{ansatz})
for the mass distribution and relative
abundances of various clusters, is consistent with the action of the RG transformation.
Another thing to notice is that Eqns~(\ref{fqgeqn1},\ref{Pswoeqn1})
are invariant under
the simultaneous
interchange of
$P_s$ and $P_w$ on the one hand, and $f_a$ and $q$ on the other.
This is to be expected, and is related to the $\delta \rightarrow -\delta$
duality of the original problem. Indeed, if we consider interchanging
$P_s$ and $P_w$, which would correspond to the $s$ type bonds
being chosen {\em weaker} than the $w$ type bonds,
then our definition of the clusters would no longer be appropriate.
In this case, it is more appropriate to think in terms of the complementary
clusters enclosing $w$-type even bonds,
and linked together by $s$-type even bonds. In this complementary
description, the roles of $f_a$ and $q$ will clearly be
interchanged. Nevertheless, the $\delta \rightarrow -\delta$ duality
implies that the equations governing the flows in these complementary
variables should have the {\em same form} as our earlier equations---in other
words, the original equations must be form-invariant under a simultaneous
interchange of $P_s$ and $P_w$, and $f_a$ and $q$.
In the rest of the article, I therefore conform without any
loss of generality to this suggestive
labeling, and restrict attention
to situations in which the $s$-type even bonds are on
average stronger than the $w$-type
even bonds, and $\delta$ is positive.

The next order of business is clearly to check our effective
model and equations against the results of Ref~\onlinecite{HymYan} for
the special $\delta = 0$ case. Let us begin by setting
$P_s \equiv P_w = P_e$, since there is no distinction at all
between the $w$-type
and $s$-type even bonds at $\delta = 0$. Next,
note that the average number of $s$-type even
bonds is given by $(1-f_a)(1-q)N_w \sum \limits_{\mu=0}^{\infty}(\mu +1)q^\mu
= N_w(1-f_a)/(1-q)$. Requiring that this equal $N_w$ gives the expected
result: $f_a = q$ at $\delta = 0$. This immediately implies that
the total number of antiferromagnetically coupled odd pairs of spin-1/2s
equals $2f_aN_w$.
Furthermore, the number of ferromagnetically coupled odd pairs is expected to
equal that of the antiferromagnetically coupled odd pairs on average.
This implies that $f_a = (2N_{ff}+N_{1f})/2N_w = (1-f_a)(1-g)$.
Moreover, the ratio $N$ of the number of spin-1s to the number of
even bonds in the system (the notation is chosen to conform with that
of Ref~\onlinecite{HymYan}) can be expressed as: $N = (2N_{11}+N_{1f})/2N_w =
(1-f_a)g$. Putting all this together, we can express $f_a$, $g$ and $q$
in terms of the single parameter $N$:
$f_a = q = (1-N)/2$, $g=2N/(1+N)$.
Furthermore, it is easy to check that Eqns~(\ref{fqgeqn1},\ref{Pswoeqn1})
at $\delta = 0$ are consistent with these relations, and
reduce to three independent equations, two for the
two functions $P_e$ and $P_o$, and one for the ratio $N$.
In fact, a little algebra shows that these reduced equations are
{\em completely equivalent}
to the equations of Ref~\onlinecite{HymYan}.

The fixed points found in Ref~\onlinecite{HymYan} can now be
translated into our language as follows: To begin with, the
RS$_1$ phase boundary at
$R>R_c(\delta = 0)$ is controlled by:
\begin{eqnarray}
&& q \; = \; f_a \; \equiv \;  (1-N)/2 \; = \; 0, \nonumber \\
&& g \; \equiv \; 2N/(1+N) \; = \; 1, \nonumber \\
&& P_s(\beta|\Gamma) \; = \; P_w(\beta|\Gamma) \; = \;
\frac{1}{\Gamma}e^{-\beta/\Gamma}, \nonumber \\
&& P_o(\beta|\Gamma) \; = \; Qe^{-Q\beta},
\label{RS1fp}
\end{eqnarray}
where $Q$ is an arbitrary constant. [Of course, this is
not really a fixed point for the probability distributions
$P_s$ and $P_w$, but becomes one if we transform to a description
in terms of the distributions of the scaled variable $\zeta = \beta/\Gamma$;
I will be sloppy about such terminology here and below,
since it is always clear what is meant from the context].
At such a fixed point, Eqn~(\ref{Nweqn1}) correctly\cite{HymYan,MonJolGol}
predicts $N_w \sim \Gamma^{-2}$.

On the other hand, the fixed points describing the $\delta =0$ Gapless Haldane phase are written as:
\begin{eqnarray}
&& q \; = \; f_a \; \equiv \; (1-N)/2 \; = \; 1/2, \nonumber \\
&& g \; \equiv \; 2N/(1+N) \; = \; 0, \nonumber \\
&& P_s(\beta|\Gamma) \; = \; P_w(\beta|\Gamma) \; = \; Pe^{-P\beta},
\nonumber \\
&& P_o(\beta|\Gamma) \; = \; Qe^{-Q\beta} \; , \; Q \; = \; 0 ,
\label{GHfp}
\end{eqnarray}
with $P$ an arbitrary constant; at any such fixed point, Eqn~(\ref{Nweqn1})
implies $N_w \sim e^{-\Gamma}$, which is consistent with the
previous results.\cite{HymYan,MonJolGol}

Finally, the
multicritical point at $R=R_c,\delta = 0$ is controlled by
\begin{eqnarray}
&& q \; = \; f_a \; \equiv \; (1-N)/2 \; = \; 1/4, \nonumber \\
&& g \; \equiv \; 2N/(1+N) \; = \; 2/3, \nonumber \\
&& P_s(\beta|\Gamma) \; = \; P_w(\beta|\Gamma) \; = \;
\frac{2}{\Gamma}e^{-2\beta/\Gamma}, \nonumber \\
&& P_o(\beta|\Gamma) \; = \; \frac{2}{\Gamma}e^{-2\beta/\Gamma},
\label{HYMCfp}
\end{eqnarray}
with Eqn~(\ref{Nweqn1}) again correctly\cite{HymYan,MonJolGol} predicting
$N_w \sim  \Gamma^{-3}$.

Let us now look for new fixed points that would correspond to the
rather unusual Griffiths phases we
expect to find in the general $\delta > 0$ case.
Consider first the (1,1) Griffiths
phase at $\delta \neq 0$: In this phase, we expect the odd bonds in the
effective model to be much weaker than both types of even bonds (this would
guarantee that the resulting pattern of singlet bonds at long
length-scales would have the topological order characteristic of the (1,1)
phase). Moreover, we expect the typical mass of clusters to scale to
zero at low-energies. Guided by such considerations, it is easy to see
that our
equations admit the following two-parameter family of fixed point
solutions that have all the `right' properties:
\begin{eqnarray}
&& f_a \; = \; 1, \nonumber \\
&& q \; = \; g \; = \; 0, \nonumber \\
&& P_s(\beta|\Gamma) \; = \; P_se^{-P_s\beta}, \nonumber \\
&& P_w(\beta|\Gamma) \; = \; P_we^{-P_w\beta}, \nonumber \\
&& P_o(\beta|\Gamma) \; = \; Qe^{-Q\beta} \; , \; Q \; = \; 0 ,
\label{11Gfp}
\end{eqnarray}
where $P_s$ and $P_w$ are two otherwise arbitrary positive constants
that obey $P_s > P_w$ (this condition matters for the stability of
these fixed points). [Note that using
these fixed point values in Eqn~(\ref{Nweqn1}) implies
$N_w \sim e^{-P_w\Gamma}$.]

Similarly, in the (2,0) Griffiths phase, we expect the $w$-type
even bonds to be much weaker than either the $s$-type even bonds
or the odd bonds. In addition, we again expect the masses of
clusters to go to zero at low-energies. Furthermore, from the
physical picture developed earlier, we expect that there are two
distinct regions in this Griffiths phase, a spin-1/2 rich region
closer to the RS$_{1/2}$ phase boundary, and a spin-1 rich region
adjacent to the RS$_1$ phase boundary.
Again, it is not hard to find the corresponding two-parameter
families of fixed points. The region closer to the RS$_{1/2}$
phase boundary is controlled by:
\begin{eqnarray}
&& f_a \; = \; g \; = \; 1, \nonumber \\
&& q \; = \; 0, \nonumber \\
&& P_s(\beta|\Gamma) \; = \; P_se^{-P_s\beta}, \nonumber \\
&& P_o(\beta|\Gamma) \; = \; P_oe^{-P_o\beta}, \nonumber \\
&& P_w(\beta|\Gamma) \; = \; Qe^{-Q\beta} \; , \; Q \; = \; 0 ,
\label{20Gfp1/2}
\end{eqnarray}
where $P_s$ and $P_o$ are otherwise arbitrary positive
constants satisfying $P_s > P_o$
(again this controls the stability of these fixed points).
[Note that using
these fixed point values in Eqn~(\ref{Nweqn1}) implies
$N_w \sim e^{-P_o\Gamma}$.]

On the other hand, the region closer to the RS$_1$ phase boundary is controlled
by:
\begin{eqnarray}
&& f_a \; = \; q \; = \; 0, \nonumber \\
&& g \; = \; 1, \nonumber \\
&& P_s(\beta|\Gamma) \; = \; P_se^{-P_s\beta}, \nonumber \\
&& P_o(\beta|\Gamma) \; = \; P_oe^{-P_o\beta}, \nonumber \\
&& P_w(\beta|\Gamma) \; = \; Qe^{-Q\beta} \; , \; Q \; = \; 0 ,
\label{20Gfp1}
\end{eqnarray}
with $P_s < P_o$ in this case (as before, this is related to the
stability of these fixed points).[Note that using
these fixed point values in Eqn~(\ref{Nweqn1}) implies
$N_w \sim e^{-P_s\Gamma}$.]

Furthermore, for the `degenerate'  crossover
case characterized by $P_s = P_o$, we have an
additional degree of freedom in the choice of $f_a$. Of course,
for any physical system, both $f_a$ and the common value, $P$, of $P_s$
and $P_w$ will be determined by complicated physics at higher energies---
the corresponding functional relationship between the two along the
cross-over line in ($R \, , \delta$) plane is clearly
outside the scope of our analysis, although we will see
later that it is possible to predict the leading behavior of both $f_a$
and $P_s$ as one approaches the multicritical point along the
crossover line. [Also, note that the non-universal value of $f_a$ along
this line of fixed points does {\em not} affect
the scaling of $N_w$, with
Eqn~(\ref{Nweqn1}) predicting $N_w \sim e^{-P\Gamma}$]

Finally, the RS$_{1/2}$ phase boundary is described by the last
specimen in our menagerie of fixed points:
\begin{eqnarray}
&& f_a \; = \; 1, \nonumber \\
&& q \; = \; 0, \nonumber \\
&& g \; = \; 1/2, \nonumber \\
&& P_s(\beta|\Gamma) \; = \; P_se^{-P_s\beta}, \nonumber \\
&& P_w(\beta|\Gamma) \; = \; P_o(\beta|\Gamma) \; = \;
\frac{1}{\Gamma}e^{-\beta/\Gamma},
\label{RS1/2fp}
\end{eqnarray}
with $P_s$ an arbitrary positive constant.[Note that using
these fixed point values in Eqn~(\ref{Nweqn1}) implies
$N_w \sim \Gamma^{-2}$, as expected.]

\section{Scaling flows and crossovers}
Naturally, the low-energy physics in any regime depends on the fixed point
controlling it via the scaling flows and crossovers in the vicinity of
the fixed point, and it is crucial to characterize these in
order to develop a complete picture of the low-energy behavior.

\subsection{Near the Multicritical point}
Let us begin in the neighborhood of the multicritical point. From
the $\delta = 0$
analysis of Ref~\onlinecite{HymYan}, we know that
the flows away from this point along the $R$ axis are controlled by
the relevant eigenvalue $\lambda_r= (\sqrt{13} - 1)/2$. A more
general stability analysis should yield another relevant eigenvalue
$\lambda_\delta$, reflecting the expected instability
of this fixed point to infinitesimal dimerization. Together, these two
exponents will then control the multicritical scaling near
this point in the usual way; in particular,
the ratio $\lambda_r/\lambda_\delta$ will determine the shape of the
RS$_{1/2}$ phase boundaries
asymptotically close to the multicritical point.

Following Refs~\onlinecite{DSF1,HymYan}, let us look for eigenperturbations
of the form
\begin{eqnarray}
&& P_s(\beta|\Gamma) \; = \; P_s^{\rm fp}(\beta|\Gamma)+
2(\epsilon_e + \epsilon_{sw})\Gamma^{\lambda-1}
(1-\frac{2\beta}{\Gamma})e^{-2\beta/\Gamma}, \nonumber \\
&& P_w(\beta|\Gamma) \; = \; P_w^{\rm fp}(\beta|\Gamma)+
2(\epsilon_e - \epsilon_{sw})\Gamma^{\lambda-1}
(1-\frac{2\beta}{\Gamma})e^{-2\beta/\Gamma}, \nonumber \\
&& P_o(\beta|\Gamma) \; = \; P_o^{\rm fp}(\beta|\Gamma)+
2\epsilon_o\Gamma^{\lambda-1}(1-\frac{2\beta}{\Gamma})
e^{-2\beta/\Gamma}, \nonumber \\
&& f_a \; = \; f_a^{\rm fp} + (\epsilon_{aq} - \epsilon_n/4)
\Gamma^{\lambda}, \nonumber \\ 
&& q \; = \; q^{\rm fp} - (\epsilon_{aq} + \epsilon_n/4)
\Gamma^{\lambda}, \nonumber \\ 
&& g \; = \; g^{\rm fp} + (4\epsilon_{n}/9 + 2\epsilon_g/3)
\Gamma^{\lambda} ,
\label{MClin1}
\end{eqnarray}
where the superscript ${\rm fp}$ denotes the values at the multicritical
fixed point. In the above, the coefficients $\epsilon$ have been chosen 
so as to separate out the effects of going away from $\delta = 0$
from the effects of changing $R$ at $\delta = 0$; thus,
$\epsilon_{aq} = \epsilon_g = \epsilon_{sw} = 0$ if $\delta$ remains zero.
Conversely, $\epsilon_n$ and $\epsilon_e$ are both expected to be
zero if $R$ remains at its multicritical value.
$\epsilon_o$ is the
only coefficient that can, in principle, get contributions both from
deviations in $R$ at $\delta = 0$, and from deviations in $\delta$ for
$R$ fixed at its multicritical value. However, one expect on physical
grounds that the distributions of the odd bonds, related as it
is to the effective couplings between the end-spins of a long segment
in the Haldane phase, is relatively insensitive to $\delta$ so long
as it is small---it is therefore reasonable to assume that $\epsilon_o$
is zero to linear order in $\delta$ at fixed $R$.
[In other words, the coefficients $\epsilon$ are related to the
perturbations in $\delta$ and $R$ as
$\epsilon_o = c_or + \dots$,
$\epsilon_e = -c_er +\dots$, and $\epsilon_{sw} =
c_{sw}\delta + \dots$, $\epsilon_{aq} = c_{aq}\delta \dots$
and $\epsilon_{n} = c_{n}r + \dots$, with all the $c$
being positive and $0(1)$, and $r \equiv R-R_c(0)$.]

This ansatz is readily seen to solve the flow equations to linear
order in the $\epsilon$, provided the following system of linear equations
is satisfied:
\begin{eqnarray}
\lambda \epsilon_g & = & -5\epsilon_g, \nonumber \\
\lambda \epsilon_n & = & 3\epsilon_g -\frac{3}{2}\epsilon_n -
\frac{3}{2}\epsilon_e + \frac{3}{2}\epsilon_o, \nonumber \\
\lambda \epsilon_e & = & -\epsilon_g -\frac{1}{2}\epsilon_n
-\frac{1}{2}\epsilon_e -\frac{1}{2}\epsilon_o, \nonumber \\
\lambda \epsilon_o & = & \epsilon_g + \epsilon_n -\epsilon_e, \nonumber \\
\lambda \epsilon_{aq} & = & -\frac{3}{2}\epsilon_{aq} +
\frac{9}{8}\epsilon_{sw} , \nonumber \\
\lambda \epsilon_{sw} & = & 2\epsilon_{aq} + \frac{1}{2}\epsilon_{sw} .
\label{MClin2}
\end{eqnarray}
Fortunately, all eigenvalues of this system can be easily
determined as follows: To begin with, we can set $\epsilon_g = \epsilon_{aq}
= \epsilon_{sw} = 0$, to obtain the three eigenvalues $-1$,
$-(1+\sqrt{13})/2$, and $\lambda_r \equiv (\sqrt{13} -1)/2$
that characterize the
$\delta = 0$ flows; the corresponding eigenvectors live entirely
in the $(\epsilon_e,\epsilon_o,\epsilon_n)$, and
correspond precisely to the eigenperturbations described
in Ref~\onlinecite{HymYan}.
Next, note that $\lambda = -5$ is clearly an eigenvalue; the corresponding
eigenvector has $\epsilon_{aq} = \epsilon_{sw} = 0$ and lives
entirely in the $(\epsilon_g, \epsilon_n, \epsilon_e, \epsilon_o)$
subspace. Finally, we have two eigenvectors that live entirely in the
$(\epsilon_{aq},\epsilon_{sw})$ subspace, with eigenvalues
$-(1+\sqrt{13})/2$ and $\lambda_\delta \equiv (\sqrt{13} -1)/2$.
Thus, we indeed have just two relevant eigenvalues, the first
$\lambda_r$ corresponding to moving along the $R$ axis at $\delta = 0$, and
the other $\lambda_\delta$ representing the effects of non-zero dimerization at
$R= R_c(\delta = 0)$. The important, and at first sight rather surprising,
feature is that these
two eigenvalues are equal! In actual fact, this
equality reflects an underlying
$S_3$ (permutation group of three elements) symmetry of the multicritical
point (corresponding to free interchange between the three phases
that meet at this point); a discussion of this point, as well as
generalizations to some closely related problems, is the subject
of a separate article.\cite{DamHus}

Now, a reasonable picture of
the full crossover from the multicritical point to any of
the phases in its vicinity is to assume that a system
very close to the multicritical point follows the multicritical
flows as the RG proceeds {\em until either}
$r\Gamma^{\lambda_r}$ {\em or} $\delta\Gamma^{\lambda_\delta}$ becomes
$O(1)$, after which the system obeys the scaling flows characteristic
of the fixed point governing the phase it is in.
This matching procedure can then be used to relate the scaling behavior of
various
quantities near the multicritical point to the two relevant eigenvalues $\lambda_r$ and $\lambda_\delta$, and I will have more to say along these
lines after we analyze the low-energy behavior of the different phases.
For now, let us ask a more basic question, namely, the shape of
the RS$_{1/2}$ phase boundaries close to the multicritical point.
It is convenient to formulate the argument in terms of
the scaling behavior of the some physical property, such as the
{\em topological order parameter}\cite{MonJolGol} $T\equiv \lim_{\Gamma
\rightarrow \infty} T(r,\delta,\Gamma)$
that characterizes the connectivity property of the (1,1) phase
(see Ref~\onlinecite{MonJolGol} for the precise definition).
Multicritical scaling, in conjunction with the results of Ref~\onlinecite{MonJolGol} for $\delta = 0$, implies that we may write
\begin{equation}
T(r,\delta,\Gamma) = \Gamma^{-(6-2\phi_M)}F(r\Gamma^{\lambda_r},
\frac{\delta^{\frac{\lambda_r}{\lambda_\delta}}}{r}),
\end{equation}
with
$\lim_{x \rightarrow -\infty}F(x,0) \sim
|x|^{\frac{6-2\phi_M}{\lambda_r}}$,
$F(0,0)$ a constant, and $\phi_M = \sqrt{5}$.
Now, if we keep $r$ fixed at some small negative value and increase $\delta$
from zero, the presence of the RS$_{1/2}$ phase boundary at which
the topological order is lost must be reflected in the the large
$\Gamma$ limit of $F$. Indeed, as we will see in the
next section, one expects $T(\Gamma)$ to scale as $\Gamma^{-(4-2\phi_{RS})}$
on the RS$_{1/2}$ phase boundary, with $\phi_{RS} = (\sqrt{5}+1)/2$. This can be consistent with
the multicritical scaling form $F$ only if $\lim_{x \rightarrow -\infty}
F(x,-y_c) \sim x^{(3-\sqrt{5})/\lambda_r}$ for some particular $y_c >0$,
implying
that $\delta^{\frac{\lambda_r}{\lambda_\delta}} = -y_c r$ is the
equation for the RS$_{1/2}$ phase boundary close to the multicritical
point. Since we have $\lambda_r = \lambda_\delta$, this means that
the phase boundary comes in {\em linearly} as shown in the schematic
phase diagram Fig~\ref{phased}. Of course, Fig~\ref{phased} contains
another ingredient, namely that the RS$_{1/2}$ line slopes {\em downwards},
ruling out reentrant behavior. To justify this needs a somewhat more
detailed consideration of the actual crossovers---while it is
possible to do this using eigenvector information and
the matching procedure outlined above, I do not pursue this further
here.

\subsection{Near the RS$_{1/2}$ and RS$_1$ phase boundaries}
The analysis in the vicinity of the RS$_{1/2}$ and RS$_1$ fixed
lines is much simpler, and the two cases are closely analogous.
Here, I discuss only the RS$_{1/2}$ phase boundary in detail,
confining myself to a brief summary of the corresponding
results in the RS$_1$ case.

In order to study small perturbations around a point on the RS$_{1/2}$
line, it is convenient to parametrize these deviations as
\begin{eqnarray}
P_s(\beta|\Gamma) & = & P_se^{-P_s\beta}[1+\delta_s(1-P_s\beta)], \nonumber \\
P_w(\beta|\Gamma) & = & \frac{e^{-\beta/\Gamma}}{\Gamma}
[1+(\delta_p-\delta_{ow})(1-\frac{\beta}{\Gamma})], \nonumber \\
P_o(\beta|\Gamma) & = & \frac{e^{-\beta/\Gamma}}{\Gamma}
[1+(\delta_p+\delta_{ow})(1-\frac{\beta}{\Gamma})], \nonumber \\
g & = & \frac{1}{2}(1+\delta_g), \nonumber \\
f_a & = & 1-\delta_a, \nonumber \\
q & = & \delta_q.
\end{eqnarray}
Requiring that this ansatz satisfy the linearized flow equations
yields a linear system of ordinary differential equations for the
functions $\delta(\Gamma)$.
Fortunately, the equations
for $\delta_a$ and $\delta_q$ are particularly simple:
\begin{eqnarray}
\frac{d\delta_a}{d\Gamma} & = & -(P_s - \frac{2}{\Gamma})\delta_a, \nonumber \\
\frac{d\delta_q}{d\Gamma} & = & -(P_s + \frac{1}{\Gamma})\delta_q +
\frac{3\delta_a}{4\Gamma}.
\end{eqnarray}
These immediately fix the $\Gamma$ dependence of $\delta_a$ and $\delta_q$
to be
\begin{eqnarray}
\delta_a & = & C_a\Gamma^2e^{-P_s\Gamma}, \nonumber \\
\delta_q & = & (\frac{C_a\Gamma^2}{4} + \frac{C_q}{\Gamma})e^{-P_s\Gamma};
\end{eqnarray}
at large $\Gamma$, both $\delta_a$ and $\delta_q$ thus decay extremely
rapidly to zero (compared to the $\Gamma^\lambda$ behavior we anticipate
for $\delta_{ow}$, $\delta_{p}$, and $\delta_g$).
Furthermore, it is easy to check that
the linearized equation for $\delta_s$ only involves
$\delta_q$ and $\delta_a$:
\begin{eqnarray}
\frac{d\delta_s}{d\Gamma} & = & -\frac{\delta_q}{\Gamma} -\frac{\delta_a}{
4\Gamma},
\end{eqnarray}
which implies that the leading large $\Gamma$ behavior of
$\delta_s$ is
\begin{eqnarray}
\delta_s(\Gamma) & \sim & {\rm const.} + \frac{C_a\Gamma}{2P_s}e^{-P_s\Gamma}.
\end{eqnarray}
The presence of the constant term in the above merely reflects the
fact that we have a whole line of RS$_{1/2}$ fixed points with
$P_s(0|\Gamma)$ varying continuously along the line (in RG terms, this
is a marginal coupling); requiring that the fixed
point value of $P_s(0|\Gamma)$ be $P_s$ at the particular
point about which we are perturbing tunes
the constant to zero, implying that $\delta_s$ also decays very
rapidly at large $\Gamma$.

This simplifies matters considerably---for clearly, we are at liberty
to set $\delta_a$, $\delta_s$, and $\delta_q$ to zero in the linearized
equations for $\delta_{ow}$, $\delta_{p}$, and $\delta_g$ in order to
determine the eigenvalues $\lambda$ that control the slow growth or
decay of these perturbations.
The corresponding linearized equations (with the
$\delta_{a/q/s}$ set to zero) read:
\begin{eqnarray}
\frac{d\delta_g}{d\Gamma} & = & \frac{1}{\Gamma}(2\delta_{ow} - 2\delta_g),
\nonumber \\
\frac{d\delta_p}{d\Gamma} & = & - \frac{\delta_p}{\Gamma}, \nonumber \\
\frac{d \delta_{ow}}{d\Gamma} & = & + \frac{\delta_{ow}}{\Gamma}
\end{eqnarray}
As in the previous section, the eigenmodes are clearly of the form
$\delta_{g/p/ow} = \epsilon_{g/p/ow}\Gamma^{\lambda}$, with
the coefficients $\epsilon$ constrained by
\begin{eqnarray}
\lambda \epsilon_p & = & -\epsilon_p, \nonumber \\
\lambda \epsilon_{ow} & = & +\epsilon_{ow}, \nonumber \\
\lambda \epsilon_{g} & = & -2\epsilon_g + 2\epsilon_{ow}.
\end{eqnarray}
The three eigenvalues are $+1$, $-1$, and $-2$, with
the respective eigenvectors proportional to $(\epsilon_p=0\,,\,\epsilon_{ow}
= 1\,,\,\epsilon_g= 2/3)$, $(\epsilon_p=1\,,\,\epsilon_{ow}
= 0\,,\,\epsilon_g= 0)$, and  $(\epsilon_p=0\,,\,\epsilon_{ow}
= 0\,,\,\epsilon_g= 1)$.

Thus, as expected, there is a single relevant eigenperturbation
on the RS$_{1/2}$ line. Clearly, this
corresponds to tuning the bare dimerization away from the critical
value $\delta = \delta_c$; from the form of the relevant eigenvector,
it is clear that this
gives rise to a slight imbalance in the strengths of the
$w$-type even bonds and the odd bonds in the effective model,
which in turn drives $g$ away from its critical value of $1/2$
(as it must, since the fixed point value of $g$ is
$1$ in the (2,0) phase, and $0$ in the (1,1) phase).
This analysis also allows us to identify the appropriate crossover
energy scale for a system close to the RS$_{1/2}$ phase boundary---
clearly, any such system will `look' critical for $\Gamma$
less than the log-energy scale
$\Gamma_{\delta_{\rm eff}} \sim \delta_{\rm eff}^{-1}$,
and will have behavior characteristic of either the (2,0) or (1,1)
phase for $\Gamma$ greater than this crossover scale (here $\delta_{\rm eff}
\equiv \delta - \delta_c)$). 
All of the above is clearly identical to the original analysis of
the spin-1/2 chain,\cite{DSF1,HyYaBhGi} and thus, the
analysis here confirms the
heuristic picture of Section~\ref{motiv}. Moreover, this also
makes it clear that scaling behavior of the topological order parameter
$T$ in the vicinity of the RS$_{1/2}$ phase boundary will be identical
to the behavior of the dimerization order parameter\cite{HyYaBhGi}
in the random spin-1/2 HAF chain in the vicinity of its RS$_{1/2}$
critical point at zero dimerization; in particular, one therefore
expect that $T(R=R_c,\delta=\delta_c,\Gamma) \sim \Gamma^{(4-2\phi_{RS})}$
on our RS$_{1/2}$ phase boundary,
with $\phi_{RS} = (\sqrt{5}+1)/2$.\cite{foot2}

The analysis near the RS$_1$ line proceeds similarly:
From the linearized flow equations, it is easy to establish
that perturbations in the values of $g$, $f_a$, and $q$ all
die away exponentially (apart from some unimportant prefactors of
powers of $\Gamma$) at large $\Gamma$. Furthermore, $P_o(\beta|\Gamma)$
also settles down to its fixed point form just as rapidly (of course,
$P_o(0|\Gamma)$ is now a marginal coupling along this line, and the particular
value $P_o$ it settles down to depends on the initial condition, analogous to
the behavior of $P_s$ in the RS$_{1/2}$ case above).
This means we are at liberty to
set all of these variables to their fixed point values in our analysis of the
much slower ($\sim \Gamma^\lambda$) growth or decay of perturbations of
$P_s$ and $P_w$ away from their common fixed point form.
Using the usual parametrization,
\begin{eqnarray}
P_s(\beta|\Gamma) & = & \frac{1}{\Gamma}e^{-\frac{\beta}{\Gamma}}
(1+(\epsilon_e+\epsilon_{sw})\Gamma^{\lambda}(1-\frac{\beta}{\Gamma})),
\nonumber \\
P_w(\beta|\Gamma) & = & \frac{1}{\Gamma}e^{-\frac{\beta}{\Gamma}}
(1+(\epsilon_e-\epsilon_{sw})\Gamma^{\lambda}(1-\frac{\beta}{\Gamma})),
\end{eqnarray}
for these slow deviations makes it clear that we again have
precisely one relevant eigenvalue $\lambda = 1$, with
the corresponding eigenvector proportional to $(\epsilon_e = 0 \, , \,
\epsilon_{sw} = 1)$ (with the irrelevant eigenvector, of eigenvalue
$\lambda = -1$, being proportional to $(\epsilon_e = 1 \, , \,
\epsilon_{sw} = 0)$). Moreover,
the relevant perturbation clearly corresponds to turning on a slight
dimerization, which introduces an imbalance in the strengths of the
$s$-type and $w$-type even bonds. Finally, the corresponding crossover
energy scale $\Gamma_{\delta}$ again behaves has $\Gamma_{\delta} \sim
\delta^{-1}$ for small $\delta$, in complete analogy with the spin-1/2
case.

\subsection{In the (1,1) Griffiths phase}
As we have seen earlier, the (1,1) Griffiths phase for $\delta \geq 0$
is described by a two-parameter family of fixed points,
with $P_s \equiv P_s(0|\Gamma$
and $P_w \equiv P_w(0|\Gamma)$ allowed to vary independently of
each other, modulo the constraint $P_s \geq  P_w$.
As expected, a formal stability analysis at any point with $P_s > P_w$
yields two marginal eigenperturbations (with $\lambda = 0$)
corresponding to these two free
parameters, with all other perturbations dying away exponentially in
$\Gamma$.\cite{foot1} However, most low energy properties in
the phase are controlled by precisely these other terms that decay
exponentially in $\Gamma$---after all, the exponential decay in the log-energy
$\Gamma$ translates to power-law behavior in the energy $\Omega$, and
it is precisely these power-laws that are the characteristic signature of
any Griffiths phase.

To analyze the low-energy flows in sufficient detail to get at this
behavior, it is useful to parametrize small perturbations away from any
$\delta > 0$ fixed point labeled as $(P_s, P_w)$ (with $P_s > P_w$) as
\begin{eqnarray}
P_s(\beta|\Gamma) & = & P_se^{-P_s\beta}(1+\epsilon_s(1-P_s\beta)),\nonumber \\
P_w(\beta|\Gamma) & = & P_we^{-P_w\beta}(1+\epsilon_w(1-P_w\beta)),\nonumber \\
P_o(\beta|\Gamma) & = & P_o(\Gamma)e^{-P_o(\Gamma)\beta}
(1+\epsilon_o(1-P_o(\Gamma)\beta)), \nonumber \\
g & = & \epsilon_g, \nonumber \\
f_a & = & 1-\epsilon_a, \nonumber \\
q & = & \epsilon_q,
\end{eqnarray}
with $P_o(\Gamma) \equiv C_oe^{-P_w\Gamma}$---this choice of
$P_o(\Gamma)$ `factors' out the $\Gamma$ dependence expected `at'
the fixed point, with $\epsilon_o$ representing a small sub-dominant
correction. Moreover, this ansatz clearly satisfies the flow equations
to leading order in the $\epsilon$ so long as the $\epsilon$ obey 
a corresponding system of ordinary differential equations that governs
their $\Gamma$ dependence.

Now, a simple analysis of the form of these equations reveals that
the perturbations $\epsilon_s$, $\epsilon_w$, and $\epsilon_o$ and their
$\Gamma$ dependence play
no role in determining the leading large $\Gamma$ asymptotics
of $\epsilon_g$, $\epsilon_a$, and $\epsilon_q$.
The $\epsilon_{s/w/o}$ can thus be set to zero as far as the analysis
of the other perturbations is concerned, and the results of this
analysis for $\epsilon_{g/a/q}$ can then be `fed back in' to
work out the leading behavior of the $\epsilon_{s/w/o}$ (this
latter step is actually quite unimportant, for it turns out
that the low-energy properties of the system depend crucially on
the behavior of the $\epsilon_{g/a/q}$, and very little on that
of $\epsilon_{s/w/o}$).
This simplifies the equations for $\epsilon_{g/a/q}$ considerably,
yielding:
\begin{eqnarray}
\frac{d\epsilon_a}{d\Gamma} & = & -(P_s - P_w) \epsilon_a, \nonumber \\
\frac{d\epsilon_g}{d\Gamma} & = & -P_w \epsilon_g + C_oe^{-P_w \Gamma},
\nonumber \\
\frac{d\epsilon_q}{d\Gamma} & = & -P_s\epsilon_q +P_w\epsilon_a,
\end{eqnarray}
where, on the right hand side of each equation, only the
terms that matter the most for the large $\Gamma$ asymptotics have
been kept.
The solution to these is relatively straightforward to write down:
\begin{eqnarray}
\epsilon_a & = & C_ae^{-(P_s-P_w)\Gamma}, \nonumber \\
\epsilon_g & = & (\Gamma C_o + C_g)e^{-P_w\Gamma}, \nonumber \\
\epsilon_q & = & C_a(e^{-(P_s-P_w)\Gamma}-C_qe^{-P_s\Gamma}),
\end{eqnarray}
where $C_a$, $C_q$, and $C_g$ are constants of integration that
depend on the initial conditions. Translating from log-energies
to the energy scale $\Omega$ and keeping only the leading power-law
in each case, we thus obtain
\begin{eqnarray}
\epsilon_a & \sim & \Omega^{P_s - P_w}, \nonumber \\
\epsilon_g & \sim & (C_g + C_o\ln(\Omega_0/\Omega))\Omega^{P_w}, \nonumber \\
\epsilon_q & \sim & \Omega^{P_s-P_w}.
\label{11asymp1}
\end{eqnarray}
With this in hand, one can now quite easily work out the leading
$\Gamma$ dependence of $\epsilon_{s/w/o}$;
however, as this will play no role in our later discussion, I refrain
from displaying the results of this analysis.

To summarize, Eqn~\ref{11asymp1} implies a rather simple picture
for the effective Hamiltonian at cutoff scale $\Omega$:
This picture is in terms of type-$a$ clusters, and
{\em $0$-mass clusters} of the
$11$, $1f$, and $ff$ types,
all connected to each other by $w$-type even bonds.
Using the leading $\Omega$ dependence of $N_w$,
$N_w \sim \Omega^{P_w}$,
derived earlier, the abundances of various types of clusters
are seen to scale as
\begin{eqnarray}
N_{a} & \sim & \Omega^{P_w}, \nonumber \\
N_{ff} & \sim & \Omega^{P_s}, \nonumber \\
N_{1f} & \sim & (C_o\ln(\frac{\Omega_0}{\Omega})+C_g)\Omega^{P_s+P_w},
\nonumber \\
N_{11} & \sim &
(C_o^2\ln^2(\frac{\Omega_0}{\Omega})+2C_gC_o\ln(\frac{\Omega_0}{\Omega})
+C_g^2)\Omega^{P_s+2P_w},
\end{eqnarray}
while the magnitudes of the exchange
couplings all obey power-law distributions $P(|J|) \sim |J|^{-1+x}$,
with $x=P_s$ for $s$-type even bonds, $x=P_w$ for $w$-type even bonds, and $x
\sim \Omega^{P_w}$ for the odd bonds.
The picture that emerges is thus very similar to the results
of the heuristic argument in Section~\ref{motiv}, with the exponents
$z_{1/2}$ and $z_{1}$ introduced 
there given in terms of $P_s$ and $P_w$ as
$z_{1/2} = P_w^{-1}$, and $z_1 = (P_s+P_w)^{-1}$ (in making
this identification, I am of course ignoring the multiplicative
logarithmic corrections, as well as the sub-dominant power-laws
predicted by the more detailed analysis here). Furthermore,
since $P_s > P_w$, $z_{1}$ is indeed less than $z_{1/2}/2$, exactly
as predicted by the earlier Griffiths arguments.

With this in hand, we can now match these low-energy flows with
our earlier results for the critical and multicritical flows to
develop a reasonably accurate picture of the full crossovers as a function
of $\Omega$ for a system in the (1,1) phase close to the multicritical
point or the RS$_{1/2}$ phase boundary.
Consider first a system in the (1,1) phase, but
very close to the RS$_{1/2}$ phase boundary (and away from the multicritical
point). In this case, the system will look critical for $\Gamma$ less
than the crossover value
$\Gamma_{\delta_{\rm eff}} \sim |\delta_{\rm eff}|^{-1}$
(with $\delta_{\rm eff} \equiv \delta - \delta_c$), and then cross
over to the flows described above at lower energies (i.e higher $\Gamma$).
For instance, both $P_o(0|\Gamma)$ and $P_w(0|\Gamma)$
will scale down as $\Gamma^{-1}$ for $\Gamma$ above this
crossover scale, while $P_s(0|\Gamma)$ will remain roughly constant.
Beyond this point, $P_o(0|\Gamma)$ will decay rapidly,
$P_o(0|\Gamma) \sim |\delta_{\rm eff}|e^{-|\delta_{\rm eff}|
(\Gamma-\Gamma_{\delta_{\rm eff}})}$, while $P_{w/s}(0|\Gamma)$ will
both remain roughly fixed at their values at the
crossover scale.
Thus, the continuously varying exponent
$z_{1/2}$ does indeed diverge as we approach the RS$_{1/2}$
phase boundary, scaling as $\delta_{\rm eff}^{-1}$, while $z_1$
is smooth across the phase boundary.
Moreover, applying a similar argument to the function $T(R,\delta,\Gamma)$
yields the scaling behavior of the topological order parameter
$T$ close to the RS$_{1/2}$ phase boundary: $T(R,\delta,\Gamma)$ is
expected to scale as $\Gamma^{-(4-2\phi_{RS})}$ for $\Gamma > \Gamma_{\delta_{\rm eff}}$, and remains roughly constant as the energy is lowered further---
as a result, $T \equiv \lim_{\Gamma \rightarrow \infty} T(R,\delta,\Gamma)$
will scale as $\Gamma_{\delta_{\rm eff}}^{-(4-2\phi_{RS})}$
implying
\begin{eqnarray}
T &\sim & |\delta_{\rm eff}|^{(4-2\phi_{RS})}
\end{eqnarray}
in the vicinity of the RS$_{1/2}$ phase boundary.

Next, consider a system in the (1,1) phase for $\delta> 0$,
but very close to the multicritical point (the behavior {\em at} $\delta = 0$
has already been discussed in Ref~\onlinecite{HymYan} and
Ref~\onlinecite{MonJolGol}). Matching the multicritical flows at
higher energies with the asymptotic behavior characteristic of the
phase tells us that
$P_{s/w/o}(0|\Gamma)$ all satisfy similar scaling forms:
$P_{s/w/o}(0|\Gamma) = \Gamma^{-1}
K_{s/w/o}(r\Gamma^{\frac{1}{\lambda_r}},\delta/r)$,
where $r = R-R_c(\delta = 0)$, and the second argument
of the scaling function is determined by the fact that
$\lambda_\delta = \lambda_r$.
Furthermore, requiring that this be consistent with the expected behavior in
the (1,1) phase immediately implies that 
$P_s \equiv \lim_{\Gamma \rightarrow \infty} P_s(0|\Gamma)$ and
$P_w \equiv \lim_{\Gamma \rightarrow \infty} P_w(0|\Gamma)$
can be written as $P_{s/w} = |r|^{\frac{1}{\lambda_r}}
\Phi_{s/w}^{-}(\delta/|r|)$ when $r < 0$ (also note that
the earlier results at $\delta=0$,\cite{HymYan,MonJolGol}
imply that $\Phi_{s/w}^{-}(0)$ must be a finite constant)

In other words, the exponents $z_{1/2}$ and $z_1$ obey the scaling
forms
\begin{eqnarray}
z_{1/2} & = & \frac{1}{|r|^{1/\lambda_r}} \Xi_{1/2}^{-}(\frac{\delta}{|r|}) \nonumber \\
z_{1}  & = & \frac{1}{|r|^{1/\lambda_r}} \Xi_{1}^{-}(\frac{\delta}{|r|}),
\label{zmultscal1}
\end{eqnarray}
where $\Xi_{1/2}^{-} = 1/\Phi^{-}_{w}$, $\Xi_1^{-} = 1/(\Phi_s^{-}+\Phi_w^{-})$,
and $r$ is assumed negative. 
Of course, in order to be consistent with our earlier analysis near
the RS$_{1/2}$ phase boundary, the scaling function
$\Phi_w^{-}$ must vanish
as $|y-y_c|$ when $y\equiv \delta/|r|$ approaches
$y_c$ (corresponding to the RS$_{1/2}$ phase boundary),
while $\Phi_s^{-}$ must go smoothly to a constant.
Moreover, the present analysis tells us something
else: Since this RS$_{1/2}$ scaling sets in roughly when $|y-y_c| < 1$,
the width in $\delta$ of the critical regime controlled
by the the RS$_{1/2}$ fixed points vanishes linearly with $r$ as one
approaches the multicritical point. Furthermore, if we
write $z_{1/2} \sim a_r|\delta-\delta_c|^{-1}$ in the vicinity of
the RS$_{1/2}$ critical line, then the scaling form derived above
implies that the amplitude $a_r$ obeys $a_r \sim |r|^{1-\frac{1}{\lambda_r}}$
as we approach the multicritical point along the RS$_{1/2}$
phase boundary.

\subsection{In the (2,0) Griffiths phase}
The situation in the (2,0) Griffiths phase is quite similar.
Except on the crossover line, a formal stability analysis again
yields two marginal eigenperturbations, corresponding to the
freedom of choice of the two `coordinates' $P_s$ and $P_o$
that parametrize the corresponding family of fixed points.\cite{foot3}
Of course, as in the (1,1) case, we need to go beyond such
a formal linear stability analysis, and describe the flows in greater
detail, to get at the low-energy physics.
As before, our task is rendered easier by the fact that the deviations of
the probability distributions
$P_s$, $P_w$, and $P_w$ from their fixed point forms play
no role in such an analysis as far as the leading
large $\Gamma$ behavior of the other
perturbations is concerned, nor do they matter much in our later calculations
of various physical quantities; they may therefore be set
to zero in our calculations.

Below, I summarize the results of such an analysis
separately for each regime
of the (2,0) phase.

\subsubsection{The spin-1/2 rich regime}
It is convenient to parametrize the deviations of $f_a$, $g$, and $q$ from their
fixed point values as
\begin{eqnarray}
f_a & = & 1-\epsilon_a, \nonumber \\
g & = & 1-\epsilon_g, \nonumber \\
q & = & \epsilon_q.
\end{eqnarray}
Using this parametrization in the
flow equations for $f_a$, $g$, and $q$ (in conjunction with the fixed point
form for the distributions $P_{s/w/o}$),
it is easy to see that the $\epsilon$ obey
\begin{eqnarray}
\frac{d\epsilon_a}{d\Gamma} & = & -(P_s - P_o)\epsilon_a, \nonumber \\
\frac{d\epsilon_g}{d\Gamma} & = & -P_o\epsilon_g + P_s\epsilon_q
+C_we^{-P_o\Gamma}, \nonumber \\
\frac{d\epsilon_q}{d\Gamma} & = & -(P_s+P_o)\epsilon_q + 2\epsilon_g\epsilon_a
C_we^{-P_o\Gamma},
\end{eqnarray}
where only terms that play a role in determining the leading large
$\Gamma$ asymptotics of the $\epsilon$ have been kept.[Note that I have
used the fixed point dependence
$P_w(0|\Gamma) = C_we^{-P_o\Gamma}$ of $P_w(0|\Gamma)$ that follows
immediately from the flow equation for $P_w$ upon using the fixed
point values for all other parameters.]
These equations immediately imply
\begin{eqnarray}
\epsilon_q & = &
C_aC_w(C_w\ln^2(\frac{\Omega_0}{\Omega})+2C_g\ln(\frac{\Omega_0}{\Omega})
+C_q)e^{-(P_s+P_o)\Gamma},\nonumber \\
\epsilon_g & = & (C_w\ln(\frac{\Omega_0}{\Omega})+C_g)e^{-P_o\Gamma},
\nonumber \\
\epsilon_a & = & C_ae^{-(P_s-P_o)\Gamma}, 
\end{eqnarray}
where $C_a$, $C_q$, and $C_g$ are all constants of integration.

This gives a rather simple picture of the low-energy
effective Hamiltonian at scale $\Omega$: The description is
again entirely in terms of type-$a$ clusters, and
{\em $0$-mass clusters} of the
$11$, $1f$, and $ff$ types,
all connected to each other by $w$-type even bonds.
Using the leading $\Omega$ dependence of $N_w$,
$N_w \sim \Omega^{P_o}$,
derived earlier, the abundances of various types of clusters
are readily seen to scale as
\begin{eqnarray}
N_{a} & \sim & \Omega^{P_o}, \nonumber \\
N_{11} & \sim & \Omega^{P_s}, \nonumber \\
N_{1f} & \sim & (C_w\ln(\frac{\Omega_0}{\Omega})+C_g)\Omega^{P_s+P_o},
\nonumber \\
N_{ff} & \sim &
(C_w^2\ln^2(\frac{\Omega_0}{\Omega})+2C_gC_w\ln(\frac{\Omega_0}{\Omega})
+C_g^2)\Omega^{P_s+2P_o},
\end{eqnarray}
while the magnitudes of the exchange
couplings all obey power-law distributions $P(|J|) \sim |J|^{-1+x}$,
with $x=P_s$ for $s$-type even bonds, $x \sim \Omega^{P_o}$ for $w$-type
even bonds, and $x = P_o$ for the odd bonds. This is clearly
very reminiscent of the heuristic picture of Section~\ref{motiv},
with the exponents
$z_{1/2}$ and $z_{1}$ introduced 
there given in terms of $P_s$ and $P_o$ as
$z_{1/2} = P_o^{-1}$, and $z_1 = P_s^{-1}$ (in making
this identification, I am of course ignoring the multiplicative
logarithmic corrections, as well as the sub-dominant power-laws
predicted by the more detailed analysis here).

Thus, as expected, this regime of the (2,0) phase is dual at low-energies
to the (1,1) phase that is separated from it by the RS$_{1/2}$
phase boundary; we can pass from one to the other by interchanging
the roles of the $w$-type even bonds and the odd bonds.
Of course, in complete analogy with our earlier analysis of
the (1,1) phase, the exponents $z_1$ and $z_{1/2}$ satisfy the multicritical
scaling form Eqn~\ref{zmultscal1}, and again, $z_{1/2}$ diverges
as $a_r|\delta-\delta_c|^{-1}$ close to the RS$_{1/2}$ phase boundary,
with the amplitude scaling as $a_r \sim |r|^{1-\frac{1}{\lambda_r}}$
for $r \equiv R-R_c$ small enough.

\subsubsection{The spin-1 rich regime}
It is convenient to parametrize the deviations of $f_a$, $g$, and $q$ from their
fixed point values as
\begin{eqnarray}
f_a & = & \epsilon_a, \nonumber \\
g & = & 1-\epsilon_g, \nonumber \\
q & = & \epsilon_q.
\end{eqnarray}
Using the fixed point form for the distributions $P_{s/w/o}$ in
the flow equations, it is easy to see that the $\epsilon$ must
satisfy
\begin{eqnarray}
\frac{d\epsilon_g}{d\Gamma} & = & -P_o\epsilon_g+P_s\epsilon_q
+\epsilon_aC_we^{-P_s\Gamma}, \nonumber \\
\frac{d\epsilon_q}{d\Gamma} & = & -(P_o+P_s)\epsilon_q +
+2\epsilon_gC_we^{-P_s\Gamma}, \nonumber \\
\frac{d\epsilon_a}{d\Gamma} & = & -(P_o-P_s)\epsilon_a + 2P_s\epsilon_g,
\end{eqnarray}
where only terms that play a role in determining the leading large
$\Gamma$ asymptotics of the $\epsilon$ have been kept.[Note that I have
used the fixed point dependence
$P_w(0|\Gamma) = C_we^{-P_s\Gamma}$ of $P_w(0|\Gamma)$ that follows
immediately from the flow equation for $P_w$ upon using the fixed
point values for all other parameters.]
These equations immediately imply
\begin{eqnarray}
\epsilon_q & = &
(C_w^2C_a\ln^2(\frac{\Omega_0}{\Omega})+2C_wC_g\ln(\frac{\Omega_0}{\Omega})
+C_q)e^{-(P_s+P_o)\Gamma},\nonumber \\
\epsilon_g & = & (C_wC_a\ln(\frac{\Omega_0}{\Omega})+C_g)e^{-P_o\Gamma},
\nonumber \\
\epsilon_a & = & C_ae^{-(P_o-P_s)\Gamma}, 
\end{eqnarray}
where $C_a$, $C_q$, and $C_g$ are all constants of integration.

Once again, this gives a rather simple picture of the low-energy
effective Hamiltonian at scale $\Omega$: As before, the description is
entirely in terms of type-$a$ clusters, and
{\em $0$-mass clusters} of the
$11$, $1f$, and $ff$ types,
all connected to each other by $w$-type even bonds.
Using the leading $\Omega$ dependence of $N_w$,
$N_w \sim \Omega^{P_s}$,
derived earlier, the abundances of various types of clusters
are readily seen to scale as
\begin{eqnarray}
N_{a} & \sim & \Omega^{P_o}, \nonumber \\
N_{11} & \sim & \Omega^{P_s}, \nonumber \\
N_{ff} & \sim &
(C_{wa}^2\ln^2(\frac{\Omega_0}{\Omega})+
2C_gC_{wa}\ln(\frac{\Omega_0}{\Omega})
+C_g^2)\Omega^{P_s+2P_o},\nonumber \\
N_{1f} & \sim & (C_{wa}\ln(\frac{\Omega_0}{\Omega})+C_g)\Omega^{P_s+P_o},
\end{eqnarray}
where we have defined $C_{wa} \equiv C_wC_a$.
The magnitudes of the exchange
couplings all obey power-law distributions $P(|J|) \sim |J|^{-1+x}$,
with $x=P_s$ for $s$-type even bonds, $x \sim \Omega^{P_o}$ for $w$-type
even bonds, and $x = P_o$ for the odd bonds.
And finally, as in the spin-1/2 rich regime, the exponents
$z_{1/2}$ and $z_{1}$ introduced in Section~\ref{motiv}
can be expressed in terms of $P_s$ and $P_o$ as
$z_{1/2} = P_o^{-1}$, and $z_1 = P_s^{-1}$ (again ignoring multiplicative
logarithmic corrections, as well as sub-dominant power-laws).

We may now match our earlier results in the vicinity of the
RS$_1$ phase boundary with this low-energy picture to obtain
the scaling of $z_1$ and $z_{1/2}$ close to the the RS$_1$ line.
In a system close to the RS$_1$ line (but not close to
the multicritical point), and for $\Gamma < \Gamma_{\delta}
\sim \delta^{-1}$, $P_{s/w}(0|\Gamma)$ will both
scale as $\Gamma^{-1}$, while $P_o(0|\Gamma)$ will stay constant.
Beyond this crossover scale, $P_{s/o}(0|\Gamma)$ will both stay
roughly constant, while $P_w(0|\Gamma)$ will fall off exponentially
with increasing $\Gamma$.
This immediately implies that $z_{1/2}$ will go smoothly
to a constant as one approaches the RS$_1$ phase boundary, 
while $z_1$ will diverge as $|\delta|^{-1}$---again, this is
consistent with the Griffiths arguments in Section~\ref{motiv}.

The situation is somewhat different close to the multicritical point,
with $\delta \neq 0$:
In this case, $P_{s/w/o}(0|\Gamma)$ all obey the scaling
form introduced earlier, $P_{s/w/o}(0|\Gamma) = \Gamma^{-1}
K_{s/w/o}(r\Gamma^{\frac{1}{\lambda_r}},\delta/r)$.
The requirement that the
$x \rightarrow +\infty$ limit\cite{foot4} of the $K_{s/w/o}(x,y)$
be consistent with the behavior expected in this regime of the
(2,0) phase immediately implies that
$P_s \equiv \lim_{\Gamma \rightarrow \infty} P_s(0|\Gamma)$ and
$P_o \equiv \lim_{\Gamma \rightarrow \infty} P_o(0|\Gamma)$
can be written as $P_{s/o} = r^{\frac{1}{\lambda_r}}
\Phi_{s/w}^{+}(\delta/r)$ when $r > 0$.
In other words, the exponents $z_{1/2}$ and $z_1$ obey the scaling
forms
\begin{eqnarray}
z_{1/2} & = & \frac{1}{|r|^{1/\lambda_r}} \Xi_{1/2}^{+}(\frac{\delta}{r}) \nonumber \\
z_{1}  & = & \frac{1}{|r|^{1/\lambda_r}} \Xi_{1}^{+}(\frac{\delta}{r}),
\label{zmultscal2}
\end{eqnarray}
where $\Xi_{1/2}^{+} = 1/\Phi^{+}_{o}$, $\Xi_1^{+} = 1/\Phi_s^{+}$,
and $r$ is assumed positive. Of course, for this to be
consistent with our earlier analysis near the RS$_1$ phase
boundary, we must have $\Phi^{+}_{o}(0)$ finite and non-zero, and
$\Phi_s^{+}(x)$ vanishing linearly with $x$ for small x.
This also tells us something more about the RS$_1$ scaling
for small $r$: Clearly, the width in $\delta$ of the RS$_1$
critical region vanishes linearly with $r$ for small $r$;
furthermore, if we write $z_1 \sim a_r |\delta|^{-1}$ near the
RS$_1$ phase boundary, then
the critical amplitude $a_r$ scales as $r^{1-\frac{1}{\lambda_r}}$
for small $r$.

\subsubsection{Along the crossover line}
The analysis of the crossover case is very similar to that in
the two regimes on either side, with only some minor
differences.
As before, the deviations of the three distributions $P_{s/w/o}$
from their fixed point form plays no role in determining the
the decay of the perturbations of $g$ and $q$ from their
fixed point values. Furthermore, it is easy to see
that the leading large $\Gamma$ behavior of these perturbations
is also independent of precisely how $f_a$ approaches its
non-universal fixed point value $f$.

Parametrizing them as
\begin{eqnarray}
g & = & 1 -\epsilon_g \nonumber \\
q & = & \epsilon_q,
\end{eqnarray}
it is easy to see that they obey the following system of equations
\begin{eqnarray}
\frac{d\epsilon_g}{d\Gamma} & = & -P\epsilon_g +P\epsilon_q+fC_we^{-P\Gamma},
\nonumber \\
\frac{d\epsilon_q}{d\Gamma} & = & -2P\epsilon_q +2(1-f)C_we^{-P\Gamma}
\epsilon_g,
\end{eqnarray}
where $P$ denotes the common fixed point value of $P_{s/o}(0|\Gamma)$,
$C_we^{-P\Gamma}$ is the fixed point value of $P_w(0|\Gamma)$, and
only kept the terms that matter the most at large $\Gamma$ have been kept.
The leading large $\Gamma$ behavior of $\epsilon_g$ and $\epsilon_q$
immediately follows:
\begin{eqnarray}
\epsilon_q & = & (f(1-f)C_w^2\Gamma^2 + 2C_wC_g(1-f)\Gamma+C_q)e^{-2P\Gamma},
\nonumber \\
\epsilon_g & = & (C_g+fC_w\Gamma)e^{-P\Gamma},
\end{eqnarray}
where $C_g$ and $C_q$ are constants of integration.
This can now be used to determine the manner in which $P_{s/w/o}$
settle down to their fixed point values, as well as follow the
approach of $f_a$ to its fixed point value. However, since none of
this matters for the leading low-energy behavior of the densities
of various clusters, I do not pursue this any further here.

These results for $\epsilon_g$ and $\epsilon_q$
give a simple picture of the low-energy
effective Hamiltonian at scale $\Omega$: As in all the other
cases, the description is
entirely in terms of type-$a$ clusters, and
{\em $0$-mass clusters} of the
$11$, $1f$, and $ff$ types,
all connected to each other by $w$-type even bonds.
Using the leading $\Omega$ dependence of $N_w$,
$N_w \sim \Omega^{P}$,
derived earlier, the abundances of various types of clusters
are readily seen to scale as
\begin{eqnarray}
N_{a} & \sim & f\Omega^{P}, \nonumber \\
N_{11} & \sim & (1-f)\Omega^{P}, \nonumber \\
N_{ff} & \sim &
(1-f)(A^2\ln^2(\frac{\Omega_0}{\Omega})+
2AC_g\ln(\frac{\Omega_0}{\Omega})
+C_g^2)\Omega^{3P},\nonumber \\
N_{1f} & \sim & 2(1-f)(A\ln(\frac{\Omega_0}{\Omega})+C_g)\Omega^{2P},
\end{eqnarray}
where the constant $A$ is defined as $A = fC_w$.
[Of course, as one moves along the crossover line
and approaches the multicritical point, $f$ will tend to the
universal multicritical value of $1/4$, and $P$ will
vanish with the exponent $1/\lambda_r$.]

\section{Physical properties}
This detailed picture of the effective Hamiltonian in various parts of the
phase diagram can be used, in principle, to obtain corresponding
information about the low-temperature ($T$) thermodynamics, and the
low-frequency, low-temperature dynamics.
Below, I focus on the low temperature specific
heat and susceptibility, as these provide us with a simple,
experimentally verifiable signature of the unusual crossover from spin-1/2
to spin-1 behavior as we tune across the crossover line in the
(2,0) phase. [Only the Griffiths phases are discussed, since the behavior
at the multicritical point and the RS$_1$ phase boundary has been analyzed in
earlier\cite{HymYan,MonJolGol} work, and the properties of the RS$_{1/2}$ 
phase boundary are identical to that of a spin-1/2 chain in its
Random Singlet state, analyzed earlier by Fisher.\cite{DSF1}]

As in Ref~\onlinecite{DSF1}, the broad
power-law distributions of the couplings in the effective Hamiltonian
make it possible for us to treat the low-temperature thermodynamics
in a relatively simple way. The basic idea is to run the RG till
the cutoff $\Omega$ is reduced to be $\Omega_T \sim T$, and recognize that
the temperature $T$ dominates over almost all exchange couplings
in this renormalized problem with cutoff $\Omega_T$.
As a result, the leading low-temperature behavior of the system can
be understood in terms of the thermodynamics of free spin-1 and
spin-1/2 objects, whose densities are obtained from our earlier results
for the abundances of various clusters in the effective Hamiltonian
at scale $\Omega_T$.

For instance, in the (1,1) Griffiths phase, the leading contribution to
the low-temperature entropy will come from the entropy of the almost
free spin-1/2 degrees of freedom that make up the type-$a$ clusters.
\begin{eqnarray}
C_{V} & = &T\ln(2)\times 2\frac{dN_a(\Omega_T)}{dT},
\end{eqnarray}
where we have set $k_B$ to one.
This implies that the leading temperature dependence of
the specific heat will be a non-universal power-law of $T$:
$C_V \sim T^{\frac{1}{z_{1/2}}}$.
These spin-1/2 degrees of freedom also provide the dominant contribution
to the zero-field susceptibility:
\begin{eqnarray}
\chi & = & \frac{g^2}{4T} \times 2N_a(\Omega_T),
\end{eqnarray}
where $g$ is the gyromagnetic ratio, and we have also set $\mu_B$,
the Bohr magneton, to $1$. The leading temperature dependence
of $\chi$ is thus $\chi \sim 1/T^{1-\frac{1}{z_{1/2}}}$

Forming the `Wilson-ratio' $W= T\chi/g^2C_V$, it is clear that the
low-temperature limit of $W$ is
\begin{eqnarray}
W & = & {\cal C}_{1/2}z_{1/2}
\end{eqnarray}
with the {\em universal} constant ${\cal C}_{1/2} = 1/(4\ln(2))$
characteristic of a {\em spin-$1/2$ system}.
Since the spin-1 degrees of freedom are always sub-dominant in
this phase, this result holds throughout, and the situation
for general $\delta \neq 0$ is thus not very different from that predicted
for the $\delta = 0$ case in earlier work.\cite{HymYan,MonJolGol}

The (2,0) Griffiths phase, on the other hand, presents some unusual
possibilities:
Clearly, the above analysis carries over unchanged to the spin-1/2 rich regime
of the (2,0) phase, and the Wilson ratio takes the same form as in
the (1,1) phase.
However, in the spin-1 rich regime, the
thermodynamics in the low-temperature limit is dominated
by type-$11$ clusters, and therefore
controlled by the exponent $z_1$ that determines
their abundance:
\begin{eqnarray}
C_{V} = T\ln(3)\times 2\frac{dN_{11}(\Omega_T)}{dT},
\end{eqnarray}
which implies that $C_V \sim  T^{\frac{1}{z_{1}}}$ in
the low temperature limit.
Similarly, the susceptibility $\chi$ may be written as
\begin{eqnarray}
\chi = \frac{2g^2}{3T}\times 2N_{11}(\Omega_T),
\end{eqnarray}
which implies that $\chi \sim 1/T^{1-\frac{1}{z_1}}$ in
the low temperature limit.
In other words, we again have power-law dependences similar to the
spin-1/2 rich regime, with the Wilson ratio 
\begin{eqnarray}
W & = & {\cal C}_{1}z_{1}
\end{eqnarray}
again proportional to the dominant dynamical exponent.
However, the constant of proportionality is now
{\em completely different}, and takes on the {\em universal} value
$2/(3\ln(3))$ characteristic of a {\em spin-1 system}.
Finally, along the crossover line, one again predicts similar
power-law dependences controlled by the common value, $z$,
of $z_1$ and $z_{1/2}$, but the Wilson ratio will now be
completely non-universal, with $W = c z$, and
$c$ varying continuously along this line.
Thus, the low temperature measurements of the Wilson ratio
at different points in the (2,0) phase provide one way of
getting at the unusual physics of this phase in which
the system looks, at low temperatures,
like a spin-1 one chain in one regime, and a spin-1/2
chain in the other.
However, since the low temperature
thermodynamics is always controlled by the dominant
dynamical exponent, a direct
experimental handle on the second dynamical exponent at any
{\em particular point} in the phase is lacking.
Of course, the first corrections to the leading power-law behavior
are also straightforward to calculate, and the experimental
data could therefore be fit to such a more detailed formula to
obtain the sub-dominant dynamical exponent at any particular point;
unfortunately, this would probably not be a particularly compelling test
of the theoretical picture presented here.

I conclude with some speculations regarding experiments that could
be directly sensitive to physics controlled by the second
dynamical exponent:
One might imagine that both exponents will leave an imprint
on some dynamical property like the inelastic neutron
scattering cross-section. However, reasoning as in
Ref~\onlinecite{MotDamHus2}, it is clear that the low-frequency
intensity of the main feature in the spectrum
(at wavevector $\pi$) will again be controlled by the
dominant exponent, with the contribution of the sub-dominant
degrees of freedom only providing a small correction to
this dominant low-frequency behavior.
While it is possible that the low-frequency intensity
in some other parts of the Brillioun zone might have
some signature of these sub-dominant
contributions, this is not at all obvious, and
remains only a tantalizing possibility for
now. Another possibility relates
to the fact that strong {\em ferromagnetic correlations} between
some pairs of widely separated spins are
present in the ground state in either Griffiths phase. Since the
occurrence of such correlations is controlled primarily by the
second, sub-dominant exponent, any signature of these in
the {\em static} structure factor at low-temperatures could also give us a way
of directly measuring this exponent.
Finally, in a regime in which the spin-1/2s dominate,
the most direct way of seeing the spin-1s would be
to use some dynamical probe that is
preferentially sensitive to spin-1 degrees of freedom, via
a `selection-rule' requiring that
$\Delta m_z = 2$ for the transitions induced (where $m_z$ is the
$z$ projection of the spin quantum number of a state), but again,
this has not yet been backed up by any specific calculations.

\section{Outlook and conclusions}
Thus, the low-energy behavior of random antiferromagnetic spin-1 chain
systems presents a particularly interesting example of Griffiths effects
if there is some dimerization in the exchange couplings on average.
Indeed, as I have demonstrated above,
Griffiths effects in such systems lead to a rather unusual
Griffiths phase---in one region of such a phase, the system looks
like a spin-1/2 chain at low energies, while it transforms itself into a
spin-1 chain in another part of the same phase.
Somewhat fortunately, this unusual low-energy behavior
may be described in detail, and some concrete experimentally
verifiable signatures
in the low temperature behavior identified, via an `almost-exact'
analytical strong-disorder renormalization group approach.
The RG approach presented here yields a detailed picture for the effective
Hamiltonian valid at low energies in various parts of the phase
diagram, and an intriguing possibility for future work is to
use this picture to identify some compelling dynamical signatures of this
interesting low-energy physics.
Of course, Griffiths phases similar to those described here will exist in
higher spin chains as well.
Given the recent results of
Ref~\onlinecite{GilKehFis}, $S=3/2$ chains are of particular interest
in this regard, especially since such Griffiths effects could be important
even at relatively low values of randomness in these systems (the
Griffiths phases of $S=3/2$ chains will be discussed in a separate\cite{DamHusRef} article).
Finally, it would be interesting to ask if similar disorder effects may exist
in two-dimensional magnets with more than one topologically
distinct phases for the pure system.

\section{Acknowledgements}
I would like to thank Daniel Fisher for valuable
advice, discussions, and encouragement throughout the course of this work,
David Huse for useful discussions, as well as collaboration on some
closely related work,\cite{DamHus} and O.~Motrunich and G.~Zar\'and
for useful discussions and helpful comments on an earlier draft.
I would also like to thank
E. Demler, B. Halperin, W. Hofstetter,
J. Lidmar, G. Refael, S. Sachdev, and R. da Silveira for useful discussions,
R. Hyman for sending a copy of the
relevant chapter of his unpublished thesis, and
NSF grant DMR-9981283 for financial support.


\begin{references}
\bibitem{foot0} Note that our effective model bears some resemblance to the
model proposed in Ref~{\protect \onlinecite{GilKehFis}} to analyze
the effects of randomness on antiferromagnetic $S=3/2$  chains, although
the physics is quite different.
\bibitem{foot2} Here, we have used the corrected value of $\phi_{RS}$,
quoted in a footnote of Ref~{\protect \onlinecite{HymYan}}, instead of the
value derived earlier in Ref~{\protect \onlinecite{HyYaBhGi}}; see also
Ref~{\protect \onlinecite{MotDamHus1}} below for a `percolation'
picture reducing this to Fisher's original result{\protect \cite{DSF2}}
for the corresponding
magnetization exponent of the random transverse field  Ising chain.
\bibitem{foot1} The Griffiths phase at $\delta = 0$, with $P_s = P_w$
constitutes a special case---at any point along this line, there is,
in addition to two marginal perturbations, a marginally relevant
perturbation (growing as a power of $\Gamma$), which simply reflects
the fact that any point in the (1,1) phase with $\delta > 0$ has
$q =0$, reflecting the extreme paucity of $s$-type even bonds
relative to the $w$-type even bonds at low enough energies, while, precisely at $\delta = 0$, there is no such distinction between the $s$-type
and $w$-type even bonds, and $q = 1/2$.
\bibitem{foot3} Along the crossover line, one has, in addition
to the two marginal perturbations generic to the (2,0) phase, another
marginally relevant perturbation; this is simply the formal
way of saying that $f_a$ is driven to either $1$ or $0$
as soon as one moves off the crossover line, while at any
point along this line, it takes some
non-universal value between these two extremes.
\bibitem{foot4} The $x \rightarrow +\infty$ limit is appropriate
because the RS$_1$ line exists only for $r> 0$.
\end{references}
\end{document}